\documentclass[12pt,preprint]{aastex}

\slugcomment{To appear in ApJ}

\shorttitle{VMR-D cloud IRAC Point Source Catalog}
\shortauthors{Strafella et al.}

\begin{document}


\title{The {\it Spitzer}-IRAC Point Source Catalog of the Vela-D Cloud }


\author{ F. Strafella\altaffilmark{1}, D. Elia \altaffilmark{2,1}, L. Campeggio\altaffilmark{1}, T. Giannini\altaffilmark{3}, D. Lorenzetti\altaffilmark{3}, M. Marengo\altaffilmark{4}, H.A. Smith\altaffilmark{5}, 
G. Fazio\altaffilmark{5}, M. De Luca\altaffilmark{3,6}, F. Massi\altaffilmark{7}} 

%
%
%


\altaffiltext{1}{Dipartimento di Fisica, Universit\`a del Salento, CP 193, 73100 Lecce, Italy, francesco.strafella,loretta.campeggio@le.infn.it}
\altaffiltext{2}{Universidade de Lisboa, Faculdade de Ci\^encias, Centro de Astronomia e Astrof\'{\i}sica, Observat\'orio Astron\'omico de Lisboa, Tapada da Ajuda 1349-018, Lisboa, Portugal, eliad@oal.ul.pt}
\altaffiltext{3}{INAF - Osservatorio Astronomico di Roma, via Frascati 33, I-00040 Monte Porzio, Italy, giannini,dloren@oa-roma.inaf.it}
\altaffiltext{4}{Department of Physics and Astronomy, Iowa State University, Ames, IA 50011}

\altaffiltext{5}{Harvard-Smithsonian Center for Astrophysics, Cambridge, MA 02138, mmarengo,hsmith@cfa.harvard.edu}
\altaffiltext{6}{LERMA-LRA, UMR 8112, CNRS, Observatoire de Paris and Ecole Normale Superieure, 24 Rue Lhomond,
75231 Paris, France, massimo.de.luca@lra.ens.fr} 
\altaffiltext{7}{INAF - Osservatorio Astrofisico di Arcetri, Largo E. Fermi 5, 50125 Firenze, Italy, fmassi@arcetri.astro.it} 

\begin{abstract}
This paper presents the observations of the Cloud D in the Vela Molecular Ridge, 
obtained with the IRAC camera onboard the {\it Spitzer Space Telescope} at the 
wavelengths $\lambda$= 3.6, 4.5, 5.8, 8.0~$\mu$m. A photometric catalog 
of point sources, covering a field of approximately 1.2 square degrees, has been extracted 
and complemented with additional available observational data in the millimeter region. 
Previous observations of the same region, obtained with the {\it Spitzer} MIPS 
camera in the photometric bands at 24$~\mu$m and 70~$\mu$m, have also been 
reconsidered to allow an estimate of the spectral slope of the sources in a wider spectral range. 
A total of  170,299 point sources, detected at the 5 $\sigma$ sensitivity 
level in at least one of the IRAC bands, have been reported in the catalog. 
There were 8796 sources for which good quality photometry was obtained 
in all four IRAC bands. For this sample, a preliminary characterization of the 
young stellar population based on the determination of spectral slope is 
discussed; combining this with diagnostics in the color-magnitude 
and color-color diagrams, the relative population of young stellar objects  
in the different evolutionary classes has been estimated and 
a total of 637 candidate YSOs have been selected. The main differences in 
their relative abundances have been highlighted 
and a brief account for their spatial distribution is given. 
The star formation rate has been also estimated and compared with the 
values derived for other star forming regions. 
Finally, an analysis of the spatial distribution of the sources by means of 
the two-point correlation function shows that the younger population, 
constituted by the Class I and flat-spectrum sources, is significantly more 
clustered than the Class II and III sources.

\end{abstract}




\section{Introduction}  \label{introduction}
The Vela Molecular Ridge (hereinafter VMR) is a cloud complex constituted 
by four giant molecular clouds \citep{Mur91}, hosting large regions of active 
star formation (hereinafter SF), in which low and intermediate mass stars are forming \citep{Lis92}. 
Its position on the Galactic plane ($272\degr \gtrsim \ell \gtrsim  260\degr$; 
~$3\degr \gtrsim b \gtrsim -3\degr$) makes this complex a unique 
laboratory for studying the SF phenomenon when the 
environmental conditions are those typical of the Galactic disk. 
This is an important point to consider in extrapolating global star 
formation parameters to the Galactic scale. Indeed, the most 
commonly studied SF regions are projected outside the 
Galactic plane, in part because of the observational need to 
minimize the source confusion. The drawback is that the 
global quantities derived for these regions could not be fully 
representative of the bulk of the SF in the Galaxy. 

Our study is focused on a particular subregion of the Vela cloud complex, 
the so-called D cloud \citep[][hereinafter VMR-D]{Mur91}, that has already 
been the subject of previous papers by our group. Former studies have
shown a rich phenomenology associated with this cloud, involving outflows
\citep{Wou99,Eli07}, jets \citep{Lor02,Gia05}, and evidences of clustered 
and isolated SF \citep{Mas00,Mas03}.
After the CO emission survey of the whole VMR carried out by \citet{Yam99},
two maps of $\approx$~1 square degree region of VMR-D were obtained with 
higher resolution by \citet{Eli07} ($^{12}$CO(1-0) and $^{13}$CO(2-1) emission lines) 
and \citet{Mas07} (mm dust continuum), aimed at studying the cloud morphology and 
properties as well as their relationships with the SF process active 
in this region \citep[see also][]{Del07}. This molecular cloud has been recently 
observed with both MIPS and IRAC cameras onboard the {\it Spitzer Space Telescope} 
and the MIPS aperture photometry has already been the subject of a previous 
work \citep{Gia07}. 

In this paper we present the analysis of the {\it Spitzer}-IRAC mosaics 
of the VMR-D cloud, covering a region which mostly overlaps the field 
observed at longer wavelengths by MIPS and also corresponding to the sky 
regions already mapped in the mm spectral range and discussed in the previous 
papers. These new IRAC observations allowed us to compile a four-band 
photometric catalog of point sources that, complemented by the 
information previously obtained at longer wavelengths, 
constitutes a unique data bank for studying the young stellar 
objects (YSOs) in this cloud. 
With this as our goal we have also revisited the photometric analysis of the 
MIPS 24~$\mu$m and 70~$\mu$m mosaics to obtain more accurate PSF 
photometry. The inclusion of the longer wavelength photometry is motivated 
by our aim to determine, as accurately as possible, the spectral slopes 
of the sources, infer their evolutionary stages, and obtain a reliable statistical description of the YSO population. 

A particularly valuable characteristic of our IRAC observations 
is that the data were collected by splitting the total integration 
time into two observing epochs 4.5 months apart, thereby enabling 
a first exploration of source variability in VMR-D. Indeed we found 
a number of sources that clearly showed brightness variations; they constitute 
a significant subsample deserving special attention, and are discussed in 
a separate paper \citep{Gia09}. 

This paper is organized as follows: after a brief 
description of the observations in \S~\ref{obse}, we give an account of 
the procedure followed for source extraction and photometry for both 
IRAC and MIPS data in \S~\ref{photom}. 
In the same section we also discuss the source selection criteria as well 
as the completeness and reliability. The final VMR-D point source catalog is 
presented in \S~\ref{thePSC}. 
In \S~\ref{discussion} we focus our attention on a subsample of 8796 sources 
(out of a total of 170299) that have been detected in all the four IRAC bands 
and for which good quality photometry was obtained. In the following we shall 
refer to this subsample as the ``working sample''.
By means of color-color and color-magnitude diagrams, we select the candidate 
YSOs obtaining a census of the YSO content of the VMR-D region. In the same 
section a brief account of the clustering properties is done. 
Our conclusions are summarized in \S~\ref{concl}.

\section{Observations} \label{obse}

The observations were carried out with the Infrared Array 
Camera \citep[IRAC;][]{Faz04} onboard the {\it Spitzer Space Telescope} 
\citep{Wer04}. The data were obtained as part of the Spitzer/IRAC Cycle 3 
GTO (PID 30335, PI G. Fazio). 
The Galactic region observed was delimited 
in longitude by $264\degr 29\arcmin \gtrsim \ell \gtrsim  263\degr 00\arcmin$ 
and in latitude by $0\degr 42\arcmin \gtrsim b \gtrsim -0\degr 50\arcmin$,  
covering about 1.2 square degrees. 
This field corresponds to a significant portion of the VMR-D cloud that has been 
mapped in four spectral bands, corresponding to the four IRAC camera detectors observing 
at the wavelengths $\lambda$ = 3.6, 4.5, 5.8, and 8.0~$\mu$m, respectively. 
The effective integration time per pixel, $ t_{exp}=20.8$~s in each band, has been 
accumulated by splitting the observations in two epochs separated by 4.5 months, 
a choice mainly dictated by the desire to monitor the observed region for 
possible variability of the sources. The same splitting of the observation time 
into two epochs has also been beneficial for a better cleaning of artifacts from 
the final mosaics.

The IRAC data were obtained by mapping the whole region, in each epoch, with 12$\times$14 tiles, each observed in two dither positions. The observation was performed in IRAC 12~s HDR mode, consisting of two consecutive integrations of 0.4 and 10.4~s exposure times. The HDR mode was chosen to limit the saturation effects around the bright regions, a problem which is clearly present in the long exposure images in the neighborhood of the brightest sources. The data were reduced using the IRACproc package (Schuster et al. 2006) to combine the individual tiles into a flux calibrated mosaic on a pixel grid of 0.8627"/pix. IRACproc is a wrapper for the Spitzer Science Center mosaic software MOPEX and provides enhanced outlier (cosmic rays) rejection.

\begin{figure}
 \epsscale{1.0}
 \plotone{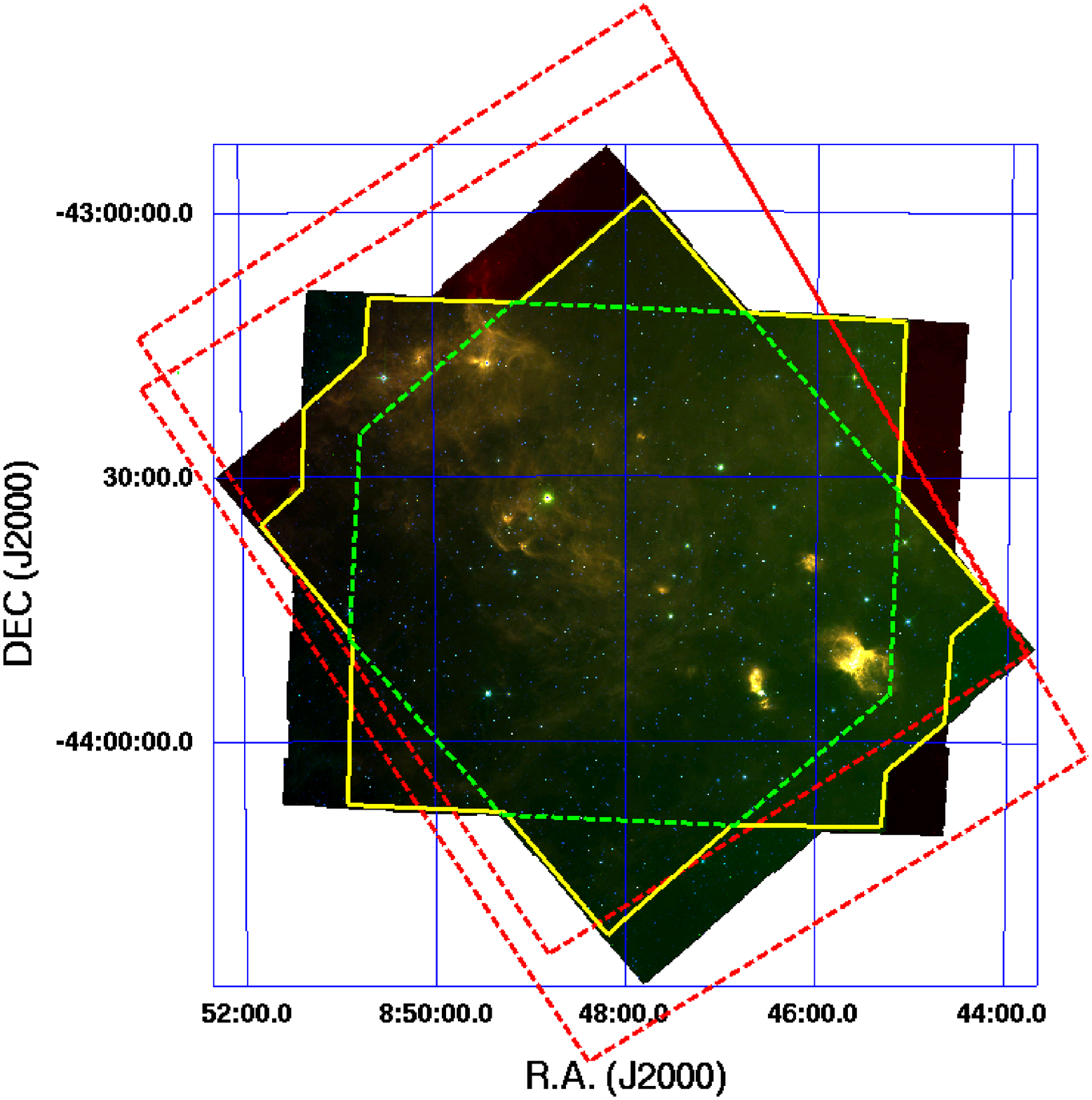}
 \caption{Three-color image of the observed field, obtained by 
 superposition of the mosaics acquired in different bands. 
 The color coding is blue, green, and red for $\lambda=$~3.6, 5.8, 
 and 8.0~$\mu$m, respectively. A yellow contour line delimits the 
 region imaged in all the four IRAC bands. 
 The geometry of the scans, obtained in two different epochs, is such 
 that only the central part of the field (dashed green line) is observed 
 twice, while more external regions are covered only once. 
 The more external red dashed lines delimit the fields observed with MIPS 
 at 24~$\mu$m (upper) and 70~$\mu$m (lower), respectively.
 \label{fig_mosaic}}
 \end{figure}

Thus for each spectral band we obtained a total of four mosaics, one long and one short exposure for each of the two epochs. The mosaics of each epoch were subsequently coadded to obtain two final mosaics in each IRAC band, one with the long ($t_{exp} = 20.8$~s) and the other with the short ($t_{exp} = 0.8$~s) total exposure time. 
However, due to the different orientation constraints in the two 
observing epochs, and to the fact that the four IRAC cameras do 
not see exactly the same fields simultaneously, the images obtained 
in the two epochs do not fully overlap. 
As a consequence, in the coadded mosaic the region corresponding 
to the full exposure time is limited by the spatial overlap of 
the mosaics obtained in the two observing epochs. The final mosaics 
therefore include a central region, observed for the total integration 
time ${\rm t_{exp}}=20.8$~s, as well as adjacent surrounding regions 
corresponding to the non-overlapping fields that have been imaged 
only in one of the two epochs for a total exposure time ${\rm t_{exp}}=10.4$~s. 
The geometry of the mapping and the different parts covered are shown 
in Figure~\ref{fig_mosaic}. 
In the following we shall consider the whole region observed in the four bands, 
delimited in Figure~\ref{fig_mosaic} by the yellow contour, to produce the 
corresponding point source catalog, recalling that the area that has been imaged 
in both epochs, included in the green dashed contour, is clearly the region 
observed with a slightly higher S/N ratio. 
In producing this catalog we also take advantage of our previous  observations 
of the VMR-D region to assess if a source falls on the mapped CO-emitting 
cloud or not. This evaluation is important and interesting in this context because 
we do not have adequate off-source observations of control fields, and the 
archival $^{12}$CO observations allow us to evaluate, at least to a first 
approximation, the possible contamination of the ON-cloud YSO population by 
OFF-cloud background/foreground objects. Operatively, we shall consider a 
source as projected on the molecular cloud, i.e. ON-cloud,  if it is 
contained within the $^{12}$CO contour line corresponding to the 
5 K~km~s$^{-1}$ level in the integrated intensity map of \citet{Eli07} which 
is shown superimposed on the IRAC 8~$\mu$m mosaic in Figure~\ref{vela_YSO+CO}. 
Clearly, this is only a necessary condition for a source to be considered 
for additional constraints aimed to discriminate bona fide ON-cloud YSOs 
from galactic/extragalactic contaminants.

\section{Source extraction and photometry} \label{photom}

To obtain the photometric information on the point sources 
detected in our mosaics, we adopted the DAOPHOT \citep{Ste87} 
photometric package. This offers a consolidated suite of 
algorithms aimed at detecting all the possible point-like 
objects whose signal is larger than a predefined threshold. 
The photometry was derived on our final (two epochs combined) mosaics. 
Single epoch photometry was however also measured for some sources, in 
order to assess the reliability of our final catalog. 
In this section we shall describe the methods used for source 
detection and photometry, as well as the selection criteria adopted 
for compiling the VMR-D point-source catalog. Completeness and 
reliability is also discussed and finally the MIPS mosaics are 
reconsidered to obtain a PSF photometry to be used as a complement 
to the IRAC catalog.

\subsection{Source detection}
Due to the large spatial extent of our mosaics and to the 
presence of extended bright nebulosity, distinct, local   
background intensity gradients are invariably present, so that 
it is not advisable to simply scan the original image 
with a given sensitivity threshold. The output of the detection 
algorithms in this case would be affected by a spatially variable 
sensitivity, depending on the different levels of the local background, 
a problem that is made even worse when strong gradients are present. 

To ameliorate this difficulty we first evaluated the point spread 
function of our mosaics by using relatively isolated, bright, 
unsaturated stars, and then performed PSF point-source subtractions 
to remove, to first order, the point-like objects. In this way we obtained  
images practically devoid of strong point sources, even if some residuals 
are clearly present. We used these images to estimate 
the background level across the whole field, a step that was done 
in each observed band by evaluating the sky value around each pixel 
in a square box of 51-pixel in size. This was done by using the 
DAOPHOT-SKY procedure that estimates the modal value after clipping 
the outliers. 
Note that the relatively large size adopted for the scanning box is 
justified by the need to include a sufficient number of sky pixels 
even in crowded images, as is the case particularly for the first two 
IRAC bands centered at $\lambda$ = 3.6~$\mu$m and $\lambda$=4.5$~\mu$m, respectively. 

Once the background in each band was obtained and 
subsequently subtracted from the original mosaics, we were left with 
four background-subtracted images in which the large scale gradients 
were strongly smoothed while the small scale noise was  
preserved. At this point, the compatibility of the noise with 
poissonian statistics is recovered by adding the original 
mean sky value back onto the background subtracted images that were then 
used in the photometry extractions. 

Source detection was then performed by scanning these final 
images with a star finding procedure which is based on a convolution 
with a wavelet-like filter whose spatial extent has been tuned to 
match the size of the point-like objects in the image. The signal 
produced by this convolution is then analyzed at different detection 
thresholds that, in our case, were taken at 5 ~$\sigma$ and 
10~$\sigma$, respectively.

\subsection{Photometry} 
Once we had the list of detections from the star finding algorithm, 
we used aperture photometry to obtain a first evaluation of the source 
magnitudes.  Although this is a relatively inaccurate method when crowding 
or background gradients are present in the image, it is nevertheless useful 
for a first estimate of source brightness, an information that 
was used later as a starting point for more accurate photometry. 

The final magnitude values were indeed obtained by means of an  
iterative PSF-fitting procedure that, in the implementation adopted 
by DAOPHOT, is based on an estimate of the instrumental PSF obtained from 
stellar images appearing relatively isolated in the frame. 
However, because the PSF is undersampled by the IRAC cameras, 
the corresponding point response function (PRF) that we obtain from our 
mosaics cannot be completely accurately determined; this produces a 
corresponding uncertainty in the subsequent fit to the stellar images. 
In spite of this difficulty, the photometry resulting from this 
method can be considered more accurate than the aperture photometry 
because the differences in fitting the PRF to the stellar images 
remain approximately balanced between positive and negative residuals. 
In this way the total flux is substantially preserved, even in presence of 
residuals, although at the cost of an increased uncertainty in the fitting 
parameters (see also $\S$~\ref{completeness} below on reliability). 
The PRF we used is actually a combination of an analytic function 
(in our case a Lorentzian showed better results in 
minimizing the residuals) and a look-up table whose aim is to compensate 
for systematic differences between the pure analytic function and the 
actual stellar images. An example of the residuals remaining after subtraction 
of the modeled sources from the IRAC mosaics is shown in Figure~\ref{fig_subtracted}, 
which illustrates this point for the two IRAC mosaics obtained at  
3.6~$\mu$m (upper panels) and the 8.0~$\mu$m (lower panels), respectively. 

Note, however, that the knowledge of the PRF also allowed us to take into account, 
and then correct for, the contribution to the flux of a source from wings in the 
brightness distribution of its neighboring sources. This advantage 
is fully exploited in crowded regions because the photometric values are 
obtained only after convergence of an iterative procedure that 
fits simultaneously groups of nearby stars instead of single objects.  

\subsection{Point source selection} \label{sourceselection}
We selected the sources to be included in our catalog by applying to 
the photometric lists obtained from the long exposure mosaics 
a sequence of logical filters based on a reasonable use of the quality 
parameters associated with the photometric procedures. 

\subsubsection{Automatically selected sources} \label{auto_selected}
Our first choice was to consider as ``good photometry sources'' only those 
detected above the 5~$\sigma$ threshold of the background noise. To easily 
distinguish faint and bright sources in the catalog, we also flagged those 
detected above the 10~$\sigma$ threshold. 

The DAOPHOT procedure for obtaining the point source photometry, 
besides evaluating the brightness, also computes two additional 
quantities: $\mathit{CHI}$ and $\mathit{SHA}$, related to the goodness of the 
fit and to the source sharpness, respectively.  
Here $\mathit{CHI}$ is evaluated as the ratio between the observed scatter 
in the fitting residuals and that expected on the basis of the 
pure image noise, and is a measure 
of the departure of the source image from the model PRF. 
The sharpness parameter $\mathit{SHA}$, instead, is computed as a difference 
between the square of the size, in pixels, of a given source and the same 
quantity computed for the fitting PRF so that it can be used, within 
certain limits, as an indicator of the source angular extent. 

At first we constrained these two parameters to be $\mathit{CHI} < 3$ and 
$ |\mathit{SHA}| < 0.7$ to represent {\it bona fide} point-like sources; we 
subsequently applied a further criterion to the flux uncertainty to 
select only sources with a flux relative error $\delta F/F < 0.5$. 
As a third step we also considered the number of iterations needed for 
the PRF fitting algorithm to converge on a given source:  to avoid the 
worst cases, mainly corresponding to faint sources in crowded fields, 
we limited selections to those sources requiring fewer than eight iterations. 
Finally, because the PRF fitting procedure also implies a recentering of the 
source position, we filtered out all those sources in which the corresponding 
best fit centroid was displaced from the original position as evaluated in the 
previous detection phase by more than 1\arcsec. 

\subsubsection{Manually dropped sources}
Despite all our efforts to select genuine point sources, the photometric lists 
still included many detections that, on visual inspection, are actually 
artifacts in the vicinity of the brightest sources \footnote{See the IRAC 
Data Handbook, ver. 3.0, avaliable at http://ssc.spitzer.caltech.edu/irac/dh/}.
This problem affects in particular the mosaics at $\lambda=$~3.6~$\mu$m
and $\lambda=$~4.5~$\mu$m, and is related to the so called {\em muxbleed} 
defect. These artifacts are recognized because they appear as a sequence of 
equally spaced pseudo-sources aligned along the same row containing 
the true bright source. Because their appearance is very much like that 
of a train of point sources, these pseudo-sources escaped all our other 
selection filters so that it was necessary to eliminate them by eye. In this 
way more than 1000 false sources were discarded.

\subsubsection{Additional sources and short exposures}  \label{additional_sources}
 Besides these false sources the previous selection criteria, applied to the long 
exposure photometry, discarded an additional, but relatively small, number of sources 
($\sim$ 1000). These are mainly located in regions affected by crowding or saturation, the 
latter case occurring especially in the neighborhood of the brightest objects. 
To compensate for this we also considered the short exposure mosaics, 
obtaining the corresponding photometry with the same procedure adopted for 
the long exposures. 
The short exposure photometry was used when, in assembling the
catalog, the long exposure photometry of a given source was rejected
due to one of the adopted selection criteria, as opposed to pixel
replacement in the mosaic or a hard cutoff between the short and long
frames.
In these cases we further checked to see if alternative, short exposure photometry 
existed that met all the catalog constraints. If it did, we included the source 
in the catalog; in this way we recovered a total of 276 sources.

We also reexamined an additional 77 rejected sources, most of which (68 cases) 
are associated with 24~$\mu$m MIPS sources, that escaped IRAC detection 
because they are either so bright they suffered saturation effects or were embedded 
in a bright nebulosity. 
For most of these sources we used the short exposure mosaics to obtain 
the photometry even though it is more uncertain.  These objects were included 
in the catalog to produce a list as complete as possible, even though they 
have been flagged as ``bad photometry'' and are not considered further in the 
analysis of the YSO population. 
However the reader can use these flux densities with their uncetainties, 
but at the same time should be aware that the corresponding quality flags 
do not satisfy all the constraints previously discussed for the automatic source 
selection.

In conclusion, we obtained four lists of sources, one for each IRAC band, 
that have been spatially delimited to include only objects lying in the sky 
region surveyed in all the four IRAC bands as delimited by the yellow 
contour line in Figure~\ref{fig_mosaic}.

\subsection{Sensitivity, Completeness and Reliability} \label{completeness}

An upper limit to the completeness of our catalog is clearly set  
by the sensitivity detection threshold adopted in the detection phase that, 
we recall, was set to 5~$\sigma$. 
This threshold corresponds to different limiting flux densities 
because the background level changes with the wavelength. We evaluated the 
limiting flux densities in each spectral band by calculating the average value 
of the pixel fluctuation in the background images obtained in the course of 
the photometric procedure. 
The same limiting flux densities are independently derived from the magnitude 
plots of Figure~\ref{hist_mag}, showing the distribution of the catalog point 
sources as a function of their magnitude in the four IRAC bands. 
From these plots it is apparent that the logarithm of the number of 
sources increases proportionally to the magnitude, until the linear 
behaviour is lost and a fall in the number of sources is clearly seen. 
The vertical line drawn in each panel denotes the largest magnitude at 
which the catalog can be considered as complete, a value that is clearly 
lower than the limiting magnitude. These values are summarized in 
Table~\ref{tab_lim_fluxes} along with their corresponding flux densities. 

As far as the reliability is concerned, we can exploit the fact that 
the VMR-D field was observed twice, allowing a check on the source detections. 
The criterion we adopt for this evaluation is to consider as reliable detections 
in a given IRAC band two possible cases: i) sources detected in both observing 
epochs, and ii) sources detected only once but showing at least another 
detection in an adjacent band. On the other hand we consider as less reliable 
detections case iii) when sources were detected only once in a given band and 
with no detection in the adjacent bands. 

To ensure that the sources we consider in estimating the reliability were 
observed twice, we restricted this analysis to the area delimited by the overlap of the 
IRAC mosaics from both epochs (dashed green line in Fig.~\ref{fig_mosaic}). 
Applying to these sources the previous criteria, we obtain the distributions 
shown in Figure~\ref{hist_mag} as a dotted (reliable) and dashed (unreliable) line. 
From these plots we estimate the reliability as a function of the magnitude by 
ratioing the number of sources satisfying the criteria i) and ii) with the 
totality of the catalog sources i)+ii)+iii) in a given band and magnitude interval. 
In this way we derive a reliability estimate that is always larger than 99\% 
in the whole magnitude range. 

All the sources observed in both epochs show flux densities that are 
 compatible within their uncertainities; the very few sources (around 0.2\%) 
 showing  a significant photometric variability have been discussed 
 in \citet{Gia09}.

   \subsection{MIPS photometry}   \label{MIPS_phot}
    As mentioned in the introduction, we also reconsidered the MIPS 
    observations at 24~$\mu$m and 70~$\mu$m of the VMR-D region covering 
    a slightly larger field of approximately 1.8 square degrees (see Fig.~\ref{fig_mosaic}). 
    These observations have already been presented by \citet{Gia07} who 
    reported the aperture photometry for 849 and 61 sources in the 
    24~$\mu$m and 70~$\mu$m band, respectively. 

    However, because aperture photometry tends to be less accurate in the presence 
    of large background gradients such as are present in the MIPS VMR-D mosaics, in 
    \citet{Gia07} the MIPS sources considered were limited to those in regions 
    where reliable aperture photometry was possible. Because PSF fitting photometry 
    is a more accurate tool, we re-analyzed the regions in the two MIPS mosaics 
    of VMR-D to obtain a revised list of sources with the corresponding PSF photometry. 
    Our aim was to obtain more accurate SEDs and then a more reliable determination 
    of the spectral slopes characterizing the YSO population. 

    The analysis of the MIPS mosaics was done with the same procedures described 
    for the IRAC images.  First, we minimized the effects of the background variations 
    by obtaining a rough subtraction of the point sources from the original image. 
    After this first step eliminated most of the point source signal, the mosaics 
    were scanned with a square box of 21 pixels on a side to produce an estimated 
    background image. This in turn was subtracted from the original mosaic producing 
    a flattening of the large scale gradients, thus allowing a subsequent point 
    source detection across the field with a much more uniform sensitivity. 
    By using a 5~$\sigma$ detection criterion we obtained two preliminary lists 
    of sources at 24~$\mu$m and 70~$\mu$m, respectively, constrained by selection 
    criteria similar to those adopted for the catalog of the IRAC sources 
    (see \S~\ref{auto_selected}). 

    A comparison between the aperture and PSF photometry methods shows that at 
    24~$\mu$m the aperture photometry tends to systematically overestimate the 
    flux by up to $\sim$50\% with respect to the PSF photometry. 
    Similarly, the flux differences at 70~$\mu$m are of the same order 
    ($\lesssim$~50\%), but they appear more erratic and without a systematic trend. 

    In the end, we obtained two more complete lists of sources measured with PSF 
    photometry, containing 1347 and 63 sources at 24~$\mu$m and 70~$\mu$m 
    respectively, and we used them to supplement the IRAC catalog. 
    Note that among the 24~$\mu$m sources we include for completeness 
    six saturated objects, corresponding to IRS 16, 17, 18, 19, 20, and 21 
    \citep[nomenclature from][]{Lis92}. In the MIPS catalog these correspond 
    to the entries ID=146, 300, 813, 814, 991, 1018, respectively,
    that had been flagged with negative 24~$\mu$m flux densities, 
    whose absolute value corresponds to the saturation limit (4~Jy). 

    We adopted angular separation as the criterion for associating the 
    24~$\mu$m and 70~$\mu$m sources, requiring that the corresponding 
    centroids be within 20\arcsec; in this way we produced a unique, merged 
    set of sources with MIPS photometry. The merged MIPS catalog is available 
    in electronic form (VMR-D\_MIPS\_catalog.txt), and it also contains 
    information on the possible associations of the millimeter  
    sources as reported in previous studies. These associations are based on 
    positional coincidences of the MIPS source centroid being within the beamsize 
    of the mm observations. 
    A short excerpt from the MIPS catalog is printed in Table~\ref{tab_MIPS_cat}. 
    The histogram of the magnitudes is presented in Figure~\ref{hist_MIPS}, and 
    it shows that the completeness flux densities are essentially the 
    same as those quoted in   \citet{Gia07}, namely $\approx 2$~mJy at 24~$\mu$m 
    and $\approx 300$~mJy at 70~$\mu$m, respectively. Some statistics of 
    the MIPS sources are presented in Table~\ref{tab_stat_MIPS}.  

   Given this new MIPS photometry, a re-examination of the color-magnitude 
   diagrams in \citet[][their Fig. 7 and 8]{Gia07}  is in order. In 
   Figure~\ref{K_24+24_70} we present the resulting K vs. K-[24] and 
   [24] vs. [24]-[70] diagrams while in Table~\ref{tab_class_MIPS-IRAC}, 2nd 
   and 3rd columns, we report the subdivision in the different classes based 
   on the spectral indexes derived by using the 2MASS K-band magnitudes as a 
   complement of the present MIPS photometry. This classification scheme is the 
   same used in \citet{Gia07} and is based on the correspondence of the colour 
   K-[24] with the spectral slope of the different classes adopted by \citet{Reb07}. 

   A comparison of the present results with those in \citet{Gia07}, based 
   on aperture photometry, is possible. We renormalized the values 
   in their Table 5 to reflect the changes in the total number of YSOs; because 
   the measurement of the flux density now has changed by up to 50\%, the numbers 
   of sources in each spectral class changes as well. We note that:

\begin{itemize} 
\item[i)] the number of MIPS sources detected now in the 24~$\mu$m band is 
     increased while at 70~$\mu$m it remains practically the same; 
\item[ii)] the percentage of ON-cloud Class I sources is now $\sim$~10\%, 
     decreased from the previously estimated value of $\sim$~30\%; 
\item[iii)] the percentage of the ON-cloud flat-spectrum sources is approximately the same, $\sim$~26\%, as compared with the previous estimate of $\sim$~29\%; 
\item[iv)] the percentage of ON-cloud Class II objects is $\sim$~62\%, an increase with respect to the previous estimate of $\sim$~37\%, while Class III sources are now $\sim$~2\% with respect to the previous value of $\sim$~4\%. 
\end{itemize}
 
Furthermore, the more accurate modeling of the point sources allows us now to 
obtain more reliable correspondences between the MIPS, IRAC, and 2MASS objects, 
alleviating in this way the problem of multiple associations. 

In the light of the significantly lower percentage we find here for Class I 
spectrum sources, we conclude that our previous suggestion for an unusual 
excess of young objects in VMR-D does not hold up based on PSF photometry, and 
in all future discussions of MIPS sources we shall refer to our current catalog.

\section{The VMR-D point source catalog}  \label{thePSC}

Once photometry was obtained we proceeded to the positional association of the 
sources detected in the different spectral bands. In this phase of band 
merging objects  detected at different IRAC wavelengths were considered to be 
associated if the relative distance of their centroids was less than 1\arcsec. 
For those cases in which, when comparing two bands, more than one source 
is found within 1\arcsec, we considered only the nearest source to be associated. 

The final VMR-D catalog was produced by including all of the sources detected 
in {\em at least} one IRAC band which fulfill all the constraints reported in 
\S~\ref{sourceselection}. An exception to this rule are the 77 sources, 
mainly corresponding to very bright objects, whose addition we justified 
in \S~\ref{additional_sources}. 
To obtain a coherent list of sources, we restricted the final catalog to the 
field observed in all the four IRAC bands, namely the region delimited by 
the yellow contour line in Figure~\ref{fig_mosaic}. 

Obviously, by construction such a catalog contains many sources lacking 
good photometry in some IRAC bands; only 8796 sources out of the full 
170,299 have accurate photometry in all four IRAC bands. However, because 
for some purposes it is useful to 
have flux densisites in all bands, we choose to give a flux density 
value in any case. This was done by reconsidering less accurate photometry previously 
discarded in the source extraction process (see \S\ref{sourceselection}) 
or, in absence of this photometry, obtaining 5~$\sigma$ upper limits for 
the source position. 
Consequently those missing flux densities in the catalog were recovered 
from the values that had been previously rejected due to the imposed constraints 
when the detection fall within a radius of 1\arcsec~from the source position. 
Note, however, that all these band filling flux densities have been flagged as 
``bad photometry'' in the final lists, in the same way the 5~$\sigma$ 
upper limits are reported in the catalog for those cases without a specific 
detection within 1\arcsec. In addition, to allow a more accurate determination 
of the spectral slopes and a more secure classification of the YSOs, 
the final catalog also considers the spatial coincidence between IRAC 
and MIPS 24~$\mu$m sources, and includes the MIPS photometry when the 
distance between centroids is within 6\arcsec. A larger distance, up to 
20\arcsec, has been adopted for MIPS sources detected only at 70~$\mu$m. 
These relatively large distances are close to the diffraction radius at 
24~$\mu$m and 70~$\mu$m, respectively, and have been adopted to include 
all the potentially associated sources. Of course, because the probability of 
the IRAC-MIPS source association decreases with the centroid distance, 
one must pay increasing attention when the angular distance is larger 
than 3\arcsec and 10\arcsec at 24$\mu$m and 70$\mu$m, respectively.
Note however that 70\% of cases have been found within 3\arcsec .
With this criterion a total of 870 MIPS 24~$\mu$m sources have been associated 
with IRAC sources; out of these, 689 are associated to a single IRAC source, 
153 are double associations, 27 are triple, and one is associated with 
four IRAC sources. To allow the reader to evaluate these associations we also 
report in the catalog the distance of the centroids for each association. 

The published version of the catalog, available in electronic form 
(VMR-D\_IRAC\_catalog.txt), summarizes the photometry arranging the information 
it in the following columns:

\begin{itemize}
\item[] Column (1): Source ID. Sources are sorted in ascending order of R.A.
\item[]Columns(2-3): Equatorial coordinates (J2000).
\item[]Columns(4-5): Galactic coordinates.
\item[]Columns(6-7): IRAC flux at 3.6~$\mu$m and its uncertainty. When the flux 
       reported is an upper limit, the uncertainty has been put to zero. 
\item[]Column(8): Quality flag, assigned as follows: the ``good photometry'' 
data are flagged with ``H'' (High) if they refer to a more than 10 $\sigma$ 
detection, and with ``L'' (Low) if the source is revealed between 5 and 10 $\sigma$, 
respectively; in addition to this, a flag ``2'' or ``1'' indicates whether 
the source lies in the sky area covered in both observing epochs, 
or only in one of them, respectively. The ``bad photometry'' data (see 
above) are flagged with ``b'', and upper limits with ``u''. 

\item[]Columns(9-17): Same as for Columns 6-8, but for 4.5~$\mu$m, 
       5.8~$\mu$m, and 8.0~$\mu$m band, respectively. 
\item[]Column(18): MIPS catalog ID number of the associated MIPS source: this is 
 reported if its centroid is within a distance of 6$\arcsec$ from the IRAC source. 
 MIPS sources detected only at 70~$\mu$m are reported when the centroid 
 is within 20\arcsec.
 Notice that, due to the different beam sizes, the association of more than one IRAC 
 source with the same MIPS source is allowed. 
\item[]Column(19): Distance in arcsec of the associated MIPS source centroid.  
\item[]Columns(20-22): MIPS flux density at 24~$\mu$m, flux uncertainty, 
 and quality flag similar to that used for IRAC flux densities but with 
 good photometry simply flagged ``G''. The symbol NaN reported for the flux density 
 denotes a position falling outside the MIPS mosaic. 
\item[]Column(23-25): The same as for column 20-22 but for the MIPS 70~$\mu$m 
 band. Note that for MIPS sources showing only the 70~$\mu$m detection (10 cases) 
 the association is reported when the IRAC sources (51 cases) are 
 within 20\arcsec~from the MIPS source. 
\item[]Columns(26-27): A flag indicating the degree of association with $^{12}$CO(1-0) 
 and $^{13}$CO(2-1) line emission, respectively, as mapped in \citet{Eli07}. 
 ``C''  indicates that the IRAC source lies on a map pixel assigned to one or 
 more gas clumps; ``A'' indicates that the IRAC source is not associated to 
 a clump, but lies inside the integrated intensity contour corresponding 
 to the 5~K~km~s$^{-1}$, adopted here as a criterion for spatial association with 
 the gas; ``O'' indicates a source lying inside the area observed in the given 
 line, but not associated with the gas according to the two previous criteria; 
 finally, ``N'' means that the IRAC source is outside the sky area covered by
 observations in the given line.
\item[]Colum(28): name of the 1.2~mm dust core, taken from the lists 
 of \citet{Mas07} and reported here when the IRAC source falls within
 the quoted core radius.
\end{itemize}

Examples of the catalog entries are given in Table~\ref{tab_IRAC_PSC} and some 
statistics of the catalog content is summarized in Table~\ref{tab_stat_global}
and Table~\ref{tab_stat_perband}.

\section{Discussion}  \label{discussion}

YSOs in SF regions can be identified by the presence of an 
infrared excess which is produced by extended circumstellar disks/envelopes 
surrounding these objects, and that participate in their pre-main sequence (PMS) 
evolution toward the main sequence. For classification purposes it is customary 
to use the $\log{\lambda F_\lambda}~ \mathrm{vs} ~\log{\lambda}$ plot to derive 
the spectral slope, $\alpha$, that is generally evaluated in the 
$2~\mu\mathrm{m}<\lambda<20~\mu$m range. 
This parameter is taken to be dependent on the evolutionary phase of the 
emitting source and its value is generally used to define four classes \citep{Lad87,And94,Gre94}: 
Class I for $\alpha \ge 0.3$, ``flat-spectrum'' for $-0.3 \le \alpha < 0.3$, 
Class II for $-1.6 \le \alpha < -0.3$, and Class III for $\alpha < -1.6 $. 
A further class, the Class 0, has been also introduced to include sources that, 
while escaping any detection at $\lambda < 10~\mu$m, show relatively strong 
submillimeter emission  \citep{And93} and should then correspond to very early 
and cold objects. This relatively simple classification scheme has been however 
criticized because of its sensitivity to inclination effects to the line of sight 
\citep[see, e.g.,][]{Rob07,Cra08}.

Alternative classification schemes based on the bolometric temperature \citep{Mye93,Che95} 
or the bolometric luminosity \citep{You05} have also been proposed in an attempt 
to minimize the influence of the system inclination to the line of sight. However 
these alternative schemes have themselves been found to be sensitive to the 
system geometry, so that we have decided to adopt the most popular classification scheme 
based on the spectral slope. This choice also allows us a more direct comparison with a 
series of studies carried out with {\it Spitzer} on Galactic star forming regions in the 
framework of the Legacy project ‘‘From Molecular Cores to Planet-forming Disks’’ 
\citep[c2d,][]{Eva03}. \defcitealias{Eva09}{E09}

The reader interested in a more detailed discussion on the effects of adopting other 
classification schemes should refer to \citet[][hereinafter E09]{Eva09}, who also 
discussed the role of the extinction corrections.

\subsection{Color analysis, and YSO selection}  \label{YSO_class}

In the following analysis we focus our attention on the 8796, out of a total of 
170,299 catalog sources, for which good photometry was obtained in all the four 
IRAC bands; i.e. the working sample already defined in \S\ref{introduction}. 
In addition to this database we also consider, when available, the 
$K$-band 2MASS photometry and the 24~$\mu$m MIPS photometry to allow a more 
accurate estimate of the spectral index in the 2~--~24~$\mu$m spectral interval. 

Our first step was the determination of this spectral index $\alpha$ from the linear 
fit of the observed $\log{\lambda F_\lambda} ~\mathrm{vs} ~\log \lambda$ points 
for all the working sample sources. 
The distributions of the $\alpha$ values obtained are shown in the two panels of 
Figure~\ref{hist_alpha}, representing the spectral indexes for ON-cloud 
(upper) and OFF-cloud (lower) sources, respectively. In the two panels the 
histograms are drawn separately for the objects in the working sample 
(solid line) and for the fraction of the candidate YSOs (dashed line) that 
have been  subsequently selected according to the criteria we shall discuss 
in the next section.

We note that the number of objects falling within the ON- and OFF-cloud 
region are quite similar (4484 against 4312). Such a fact derives from the 
similarity of both areas (0.57 against 0.60 squared deg) and indicates that 
the overall population of both regions is dominated by photospheres and 
background/foreground objects, while the YSO population possibly represents a 
negligible contribution. This latter is a common feature already found in 
surveys of other star forming regions \citep[e.g.][]{Eva09,Reb10,Gui09}. 
As a consequence, stringent criteria have to be exploited  to select the 
candidate YSOs.

\subsubsection{Source selection} \label{source_selection}
The ability of the IRAC colors to identify YSOs has been studied 
by \citet{Har06,Har07} in the Serpens cloud and by \citet{Gut09} 
in a large number of YSO clusters, and we use parts of both approaches 
here, augmented by criteria identifying AGB contaminants from \citet{Mar08}. 
After considering the locus of the 
normal reddened stars and the non-negligible extragalactic contamination, 
these authors suggested two statistical criteria to select candidate YSOs in the 
[8]~vs~([4.5]-[8]) col-mag diagram, namely: 
i) ([4.5]-[8])~$>$~0.5 and ii) [8]~$<$~14-([4.5]-[8]). 
This first criterion excludes the stellar component while the second, designed 
to statistically cut the 95\% of the extragalactic objects, has been tuned 
by studying the extragalactic component observed with IRAC in the 
Elais N1 SWIRE region. 

Applying these criteria to the VMR-D cloud we select 595 YSO candidates 
ON cloud and 244 OFF-cloud, respectively. Following \citet{Har06,Har07}, to 
include other potential 
colder candidates that could have been escaped in this diagram, we also 
considered further diagrams based on longer wavelength photometry, i.e. 
the [24]~vs~([8.0]-[24]) and [24]~vs~([4.5]-[8.0]) diagrams. 
Adopting the appropriate selection criteria 
we included 25 more sources ON and 17 OFF-cloud. 
The YSO candidates selected are shown in Figure~\ref{col_mag_col_ON} (ON-cloud) 
and  \ref{col_mag_col_OFF} (OFF-cloud) as open circles 
and grey crosses, representing objects showing 
Class I/flat and Class II/III SEDs, respectively. 

Also indicated on these plots are the original selection criteria 
from the \citet{Har06,Har07} study of the Serpens cloud. However, note that the
Serpens cloud (260 pc) is much closer than the VMR-D cloud \citep[700~pc, see][]{Lis92}. 
To reflect this, we have also indicated in the figure the region into which 
a candidate YSO in Serpens would move if displaced to the VMR-D distance. 
The region between the two inclined parallel lines can 
then be regarded as potentially containing both candidate YSOs and 
extragalactic contaminants meaning that objects in this region 
should be more accurately studied to ascertain their nature. 
Here we conservatively consider only sources delimited by the original 
criteria represented by the dashed lines in Figure~\ref{col_mag_col_ON} 
and \ref{col_mag_col_OFF}. 

In the same figure the two color ([3.6]-[4.5]) vs ([5.8]-[8.0]) 
diagram is also shown in the upper right panel with the same meaning of the 
symbols, and with a square box that is drawn to show the approximate locus 
of the Class II sources according to \citet{All04}. The irregular area 
delimited by the dot-dashed line is the locus of the AGB stars in 
this diagram, according to \citet{Mar08}.
For comparison, in a separate Figure~\ref{col_mag_col_OFF}, we show the same 
diagrams drawn for the OFF-cloud sources. 

With the aim to obtain a list of candidate YSOs as clean as possible 
from contaminants we also adopted further criteria recently discussed 
and summarized by \citet{Gut09}. These have been derived for identifying 
as likely contaminants those sources satisfying a series of constraints 
flagging star-forming galaxies and AGN, as well as shock 
excited knots that can mimic point sources in star forming environments. 
These are based on color-color diagrams as [4.5]-[5.8]~vs~[5.8]-[8.0], 
[3.6]-[5.8]~vs~[4.5]-[8.0], and [3.6]-[4.5]~vs~[4.5]-[5.8], as well as on 
the color-magnitude [4.5] vs [4.5]-[8.0] diagram. 

In Figure~\ref{col_mag_col_ONexclu} we present these diagrams (displayed 
for simplicity only ON-cloud) for the sources emerging from the previous 
selection step that are reported with crosses and circles in Figure~\ref{col_mag_col_ON}. 
Considering the constraints in \citet{Gut09} we flagged 64 sources, 42 ON 
(shown as open squares) and 22 OFF-cloud, respectively, that have then been 
eliminated from our list of candidate YSOs. 

As a further check we also applied to this list the IRAC based criteria 
of \citet{Mar08} for finding AGB contaminants, identifying in this way 
115 sources ON-cloud (112 showing a Class III SED) and 
65 OFF-cloud (62 showing a Class III SED). Also these sources have 
been eliminated from the final list. 

At the end of this process we remain with 637 IRAC selected candidate YSO sources,  
463 ON and 174 OFF-cloud that we assign to a specific class simply considering their 
spectral slope as derived considering all the available data in the 2~--~24~$\mu$m 
spectral interval. 
These sources have been used to draw the dash-dotted histogram 
in Figure~\ref{hist_alpha}. The corresponding distribution in Classes of 
these candidate YSOs is shown in 4th and 5th column of Table~\ref{tab_class_MIPS-IRAC}. 

\subsubsection{Comparison with other Star Forming Regions}

The analysis presented above allowed us to classify the VMR-D young population 
and it is interesting to compare our results to those obtained by {\it Spitzer} 
for other SF regions. These have been summarized and analyzed in a consistent 
way by \citetalias{Eva09} and we adopted it for the VMR-D comparison. 
However, because our method for selecting the YSOs is not exactly 
the same as that of \citetalias{Eva09}, before to compare our results 
it is desirable to evaluate the effects produced by applying our 
pipeline at least to one of the five clouds considered by \citetalias{Eva09}. 
The Serpens cloud is probably the most appropriate for this check: it is 
projected near the Galactic plane and has been scrutinized with accurate 
spectroscopic follow-up aimed to identify AGB contaminants \citep{Oli09}. 
For this cloud we do not attempted to select candidate YSOs out of 
the full catalog, delivered by the c2d legacy project and made available at the 
{\it Spitzer} Science Center, because our selection criteria are similar 
to those used to obtain the c2d YSO list. The difference in our approach 
is mainly in the identification of the contaminants so that we used the 
Serpens YSO catalog to analyze with our procedure the 262 sources listed. 
Applying our pipeline, based on the \citet{Gut09} criteria, to this list we 
exclude a total of 14 sources. None of these is in the list of 
the 78 candidate YSOs scrutinized by \citet{Oli09} and their distribution 
in classes is: two sources in Class I, two are flat-spectrum, eight in 
Class II, and two in Class III. 
Despite we exclude these 14 sources out of the 262 in the Serpens 
YSO catalog, we note that after their subtraction from the numbers quoted in 
\citetalias{Eva09} and reported in Table~\ref{tab_stat_YSO}, the relative 
distribution in classes remains practically the same and also the 
star formation rate is not significantly affected. In the limit of this 
indication we can compare in Table~\ref{tab_stat_YSO} the results obtained for 
VMR-D with those provided by \citetalias{Eva09}, noting that in the numbers  
quoted for the YSOs in Serpens the sources identified as AGB by \citet{Oli09} 
have been already subtracted. 

In this table the rows 1--3 consider the sources selected by means of 
the \citet{Har07} criteria and shown in Fig.~\ref{col_mag_col_ON} and 
\ref{col_mag_col_OFF}. 
The subsequent rows 4--6 report the results obtained after filtering the 
sources in the first three rows with more stringent criteria for 
eliminating extragalactic and galactic contaminants \citep{Gut09,Mar08} 
as illustrated in Fig.~\ref{col_mag_col_ONexclu}. 
Further three rows, namely 7--9, are used to present the results obtained after 
applying to the first three rows the correction for contaminants suggested by 
\citet{Oli09} on the basis of their spectroscopic investigations of the 
Serpens YSO population selected with the \citet{Har07} criteria.

Indeed, because Serpens is a SF region quite close to Galactic plane 
($l \sim$ 5$^{\circ}$), it is reasonable to assume that its level 
of contamination of AGB and giant stars is comparable to that of VMR-D. The sample 
of YSOs in Serpens is contaminated by background giants to a level of about 25\% so that, 
by applying the same correction to our VMR-D ON and OFF-cloud sample, we obtain 
the statistics given in Table~\ref{tab_stat_YSO}, lines 7--9. Note that in this case 
we do not report the distribution in classes because the 25\% contamination is 
referred to the total number of YSOs. 

In the subsequent lines 10--14 we also report for comparison the values of the 
corresponding quantities quoted by \citetalias{Eva09} 
for ChaII, Lupus, Perseus, Serpens, and Ophiucus. 
In each line we give the investigated area, the adopted cloud distance, 
the total number of YSOs found, the surface density per solid angle and per 
square parsec, the number of objects of type Class I, flat-spectrum, Class II, 
and Class III, respectively, and finally the derived star formation rate. 

First of all we recall that, differently from these other SF regions, VMR-D is 
located in the Galactic plane and it is also from 2 to 5 times farther away. 
As already mentioned, the latter circumstance imposes a larger value 
for the limiting luminosity (typically 0.1 L$_{\sun}$ instead of, e.g., 0.004 L$_{\sun}$ 
for the Taurus) resulting in significant contamination from stars; conversely, 
it offers the opportunity to study the SF process where the mass reservoir is the 
largest available. 

As a rough first approximation, we consider the OFF-cloud statistics as representative 
of the average population of the Galactic plane and, as such, it could be subtracted 
from the ON-cloud statistics after weighting for the different areas, in this way 
estimating the uncontaminated young population of VMR-D. By doing this we obtain 
the lines 3, 6, and 9, of Table~\ref{tab_stat_YSO}, with the caution that the OFF-cloud region 
could be active in SF as suggested by the the histograms in Figure~\ref{hist_alpha}, 
albeit at a lower level. Because of this and of the larger luminosity limit, the numbers 
reported here for the YSOs should actually be considered lower limits. However, taken at 
their face value, they suggest an excess of Class III ($\sim$22\%) and a 
deficit of Class II objects ($\sim$48\%) with respect to the other SF regions. 
The cause of the apparent excess of Class III might be either evolution - the Vela 
cloud is older than these others - or due to uncorrected contamination of AGB stars 
\citepalias[see][]{Eva09}.

Considering the statistics corrected for extragalactic and galactic contaminants, 
reported in the 4th and 6th lines, make the distribution in classes of the YSOs in 
VMR-D resemble more that seen in Serpens and Ophiuchus rather than in other 
SF regions. Differences actually exist in the Class II/III relative 
ratios, but they probably stem from an incomplete correction since VMR-D is more embedded 
in the plane than is Serpens. If we try to account for such discrepancies, we can 
reasonably rule out evolutionary effects since the relatively large fraction 
of Class III sources in VMR-D does not reflect a particularly small fraction 
of Class I and flat-spectrum sources that instead appear relatively numerous. 
Moreover, if we calculate the SF rate per square parsec, by assuming a mean mass of 
1 $M_{\sun}$, a fraction of binary of 0.5 and a period of 2 Myr for SF \citepalias{Eva09}, 
we obtain for the VMR-D SF rate $\approx$ 2.5--4.1 M$_{\sun}$Myr$^{-1}$pc$^{-2}$, 
a range of values that is more consistent with that of the younger clouds like Serpens, 
Ophiucus, and Perseus, than it is for the smaller corresponding values obtained for 
Lupus and Chamaleon. 

It is also noteworthy that adopting the corrections suggested by \citet{Oli09} 
we obtain (Tab.~\ref{tab_stat_YSO}, lines 6--8) very similar results to those obtained 
after considering the set of criteria discussed in \S~\ref{source_selection} to exclude 
likely contaminants (lines 4--6). This adds some confidence in the present results.

Given these observational indications it seems plausible to consider the VMR-D age 
as being in the range 1--2 Myr, in good agreement with the dynamical time estimate 
of 1.5 Myr derived by \cite{Eli07} for VMR-D and comparable to the estimates for 
Perseus \citep{Pal00} and Serpens \citep{Dju06}. Within the previous assumptions, and 
given the solid angle subtended by our ON-cloud region, it is easy to estimate that 
approximately 200--300 M$_\sun$ Myr$^{-1}$ are converted into stars in the investigated 
region. Because the cloud mass has been estimated at $1.5\times10^4$ M$_\sun$ \citep{Eli07}, we 
can also derive the depletion time for the cloud, that is $t_{depl}\approx$~40--70 Myr, 
as well as the SF efficiency given by M$_{\mathrm{YSO}}$/(M$_{\mathrm{YSO}}+$M$_{cloud}$)$\approx$~0.02--0.03. 
However, the uncertainty associated to the estimate of the cloud mass is at least 
$\Delta M/M\sim 50\%$ so that these quantities are correspondingly uncertain.

\subsection{IRAC vs. MIPS-based classification and final lists of YSO}  \label{comp_MIPS_IRAC}

We have seen that the IRAC data are an essential, additional ingredient 
of YSOs classification, and by including them we have been able to reconcile 
(as indeed was anticipated by \citet{Gia07}) the estimated excess of Class I 
objects previously derived using only MIPS aperture photometry.

In this section we want to briefly discuss the young population census, 
as obtained by means of the present IRAC photometry complemented with 
2MASS and MIPS data, with the results that can be derived using only 
2MASS and MIPS photometry.  The aim is to assess the ability of the MIPS 
photometry to correctly classify YSOs when no IRAC data are available, as 
is the case for the MIPS regions without overlap with the IRAC mosaics 
(see Fig.~\ref{fig_mosaic}).

The classification method used in these two cases is different: 
in the first case the 2MASS/IRAC/MIPS photometric data are used to 
evaluate the spectral slope $\alpha$ and to separate candidate 
YSOs from contaminants, while in the second case only 2MASS/MIPS data 
are used to exploit the K$_s$ vs. K$_s$-[24] diagram as in \citet{Gia07}. 
The first method is clearly the most appropriate involving more photometric 
information (up to six wavelengths); however, being the second method the 
alternative choice when MIPS observations are the only available, it is 
interesting to check if differences arise in selecting the YSO population 
with these two methods. 
The two lists of candidate YSOs obtained with these two approaches are 
reported in the corresponding electronic catalogs: VMR-D\_IRAC\_YSO.txt and 
VMR-D\_MIPS\_YSO.txt. Typical entries for both are reported in 
Table~\ref{tab_IRAC_YSO} and \ref{tab_MIPS_YSO}, respectively. 

In discussing the differences emerging between these two approaches 
we shall limit ourselves to consider only the corresponding lists of 
Class I objects; similar comparisons for the other Classes (flat, II or III) 
follow the same line of thinking.  Before proceeding, we reiterate that the 
census of the different Classes in this paper is based on the working sample, 
i.e. is limited to those sources with valid fluxes in all four IRAC bands. 
This clearly excludes the sources with an upper limit, a large flux uncertainty, 
or even a not strictly point-like shape in at least one IRAC band. 
These objects, although not considered here, are included in the general 
catalog and could represent interesting targets for future investigations. 

The 49 sources selected as Class I objects by scrutinizing the 
IRAC catalog working sample are listed in Table~\ref{tab_IRAC_YSO_CI}. 
Along with the source ID, coordinates and spectral index, we also report 
the presence of a counterpart in other bands: K$_s$-2MASS in column~5 
and MIPS bands in columns~6~to~8.
This information is needed to compare the two methods since 
the MIPS classification is based on K$_s$-2MASS and 24~$\mu$m magnitudes, and  
only IRAC sources associated with both these ancillary data are subject to both 
IRAC and MIPS classification. 
By looking at Table~\ref{tab_IRAC_YSO_CI} (column 9), we see that only five 
sources out of the 49 entries were classified with MIPS and all of these are 
Class I (flag I). While this is the expected result when the four IRAC 
points are reasonably well aligned with the two extreme spectral points at 
K$_s$ and 24~$\mu$m, it is however noteworthy that in the relatively few 
cases in which both methods can be applied they give the same 
classification. 
In columns 10 and 11 we also indicate whether the source is associated with 
gas and dust emission. In this respect, the genuine nature of the selected 
Class I sources is testified by the fact that 44 out of the 49 sources 
are coincident with gas clumps or peaks. The remaining 5 sources are located 
in regions not covered by the $^{12}$CO map (flag N). 
Finally, the last column indicates whether we find that the source is also 
IRAC variable on a timescale of 4.5 months \citep{Gia09}; 
we identify six cases of variability.

Table~\ref{tab_MIPS_YSO_CI} is complementary to Table~\ref{tab_IRAC_YSO_CI}. 
It provides the list of all the 20 sources classified as Class I based on 
K$_s$-2MASS and 24~$\mu$m photometry. Besides the K$_s$-24 index (column~4) 
information is also given about the presence of a 70 $\mu$m counterpart 
(column~5). The association with an IRAC counterpart is indicated in 
column~6 while column~7 reports the quality flags of the detections in 
the four IRAC bands.  Since this classification is determined by MIPS, 
there are only seven sources (see column~8) contained in the IRAC working 
sample, while 11 further sources are associated with IRAC objects 
lacking good photometry in one or more bands.  The remaining two 
sources do not have a counterpart in the IRAC catalog (see the notes to 
Table~\ref{tab_MIPS_YSO_CI}). 

Five objects in this table are clearly in common with 
Table~\ref{tab_MIPS_YSO_CI} but there are two more cases included in 
the working sample and classified as flat-spectrum sources when we 
derive the spectral slope exploiting the IRAC data. This suggests that 
the classification independently provided by K$_s$-2MASS/MIPS and 
K$_s$-2MASS/IRAC/MIPS is not completely equivalent even if the two 
methods seem to be in agreement in five out of the seven common cases 
encountered. 
Again, most of these sources are located in gas clumps. Note also 
that Table~\ref{tab_MIPS_YSO_CI} contains many sources whose IRAC flags 
do not fulfill our compelling requirements (only H1 and H2 quality flags). 
Although they are not members of the working sample, they are nonetheless 
good candidates for very young protostars. 

Finally, for the sake of completeness, we present in Table~\ref{tab_MIPS_remark} a 
list of MIPS selected sources that, for various reasons, can be considered 
as particularly red or cold.  When observed in both MIPS bands, these objects have 
a color [24]-[70] $>$ 6.3, a condition we imposed to select the coldest (T $<$ 50K) 
sources (see Figure~\ref{K_24+24_70}, right panel). Other sources listed in 
Table~\ref{tab_MIPS_remark} are: {\it i}) the 70 $\mu$m sources with point-like appearance 
but without a 24~$\mu$m counterpart;  {\it ii}) the objects saturated at 24~$\mu$m, but already 
classified as IRAS Class I objects (Liseau et al. 1992). Note that increasing the angular 
distance of the association, reported in column 7, the uncertainty also increases 
so that further analysis is required to confirm a genuine source association.

\subsection{Clustering}
The spatial distribution of the young stars provides useful information about 
SF mechanisms in the cloud, since YSOs do not move very far from their birthplaces 
during the PMS evolution. Much of the accumulated observational evidence suggests 
that stars prefer to form in groups, and in particular, that the two-point correlation 
function of young stars shows a different behaviour at small and large spatial scales, 
respectively \citep[see, e.g.,][]{Sim97}. More recently \citet{Gut09} analyzed the 
spatial distribution of the YSOs in 36 clusters by means of both the nearest neighbor 
and the minimum spanning tree method. A critical length, in the range $\sim$ 0.1--0.5~pc, 
could be identified in the cumulative distribution of the source spacings indicating 
differences in the slope at small and large spatial scales. 
We use here the function $w(\theta)$ to quantify the degree of the YSOs clustering 
in the VMR-D cloud: 
\begin{equation}
  w(\theta)=\frac{N_p(\theta)}{N_0(\theta)}-1
\end{equation}
where $N_p(\theta)$ is the average number of pairs with angular separation between 
$\theta$ and $\theta + \Delta\theta$ around each object, and $N_0(\theta)$ is the 
same number expected for a random distribution of the same sources.
For SF regions, this function for PMS objects typically shows a power law behaviour 
$w(\theta)\propto \theta^{-\gamma}$, but with clearly different exponents for small 
($\gamma\approx 2$) and large ($\gamma\approx 0.5$) spatial scales, respectively 
\citep[e.g.][]{Lei93,Lar95,Sim97}. 
This fact has been interpreted as an indication of a critical scale that separates 
the regime of binary and multiple stars, at smaller scales, from the hierarchical 
clustering dominating at larger scales. \citet{Lar95} also suggested that this scale 
could represent the Jeans length of the molecular cloud, although this interpretation 
encounters difficulties in explaining the different length scales observed in Taurus, 
Ophiuchus and Trapezium clusters \citep{Sim97}.

\begin{figure}
 \epsscale{1.1}
 \plotone{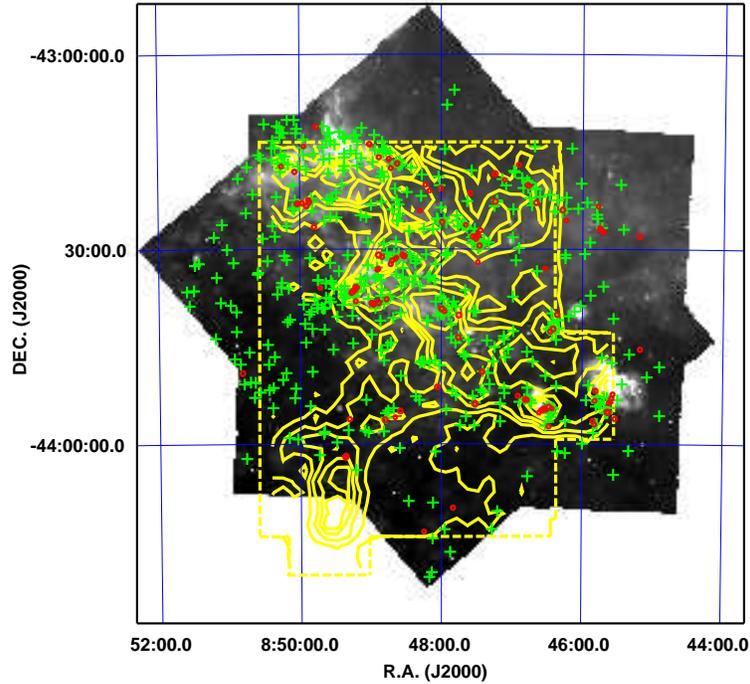}
 \caption{ The $^{12}$CO contours from \citet{Eli07} are reported (yellow line) superposed 
 to the IRAC 8.0~$\mu$m mosaic, starting at 5~K~km~s$^{-1}$ in steps of 15~K~(km~s$^{-1}$). 
 The yellow dashed line delimits the region observed in $^{12}$CO.
 The spatial distribution of the Class I/flat (red circles) 
 and the Class II/III (green crosses) is also shown. The reported sources are those recognized 
 as {\it bona fide} YSOs by using the selection criteria of \citet{Har07} and the exclusion 
 criteria of \citet{Gut09} discussed in \S~\ref{source_selection} and 
 illustrated in Figure~\ref{col_mag_col_ON}, \ref{col_mag_col_OFF}, and \ref{col_mag_col_ONexclu}. }
 \label{vela_YSO+CO}  
 \end{figure}

\begin{figure}
 \epsscale{.70}
 \plotone{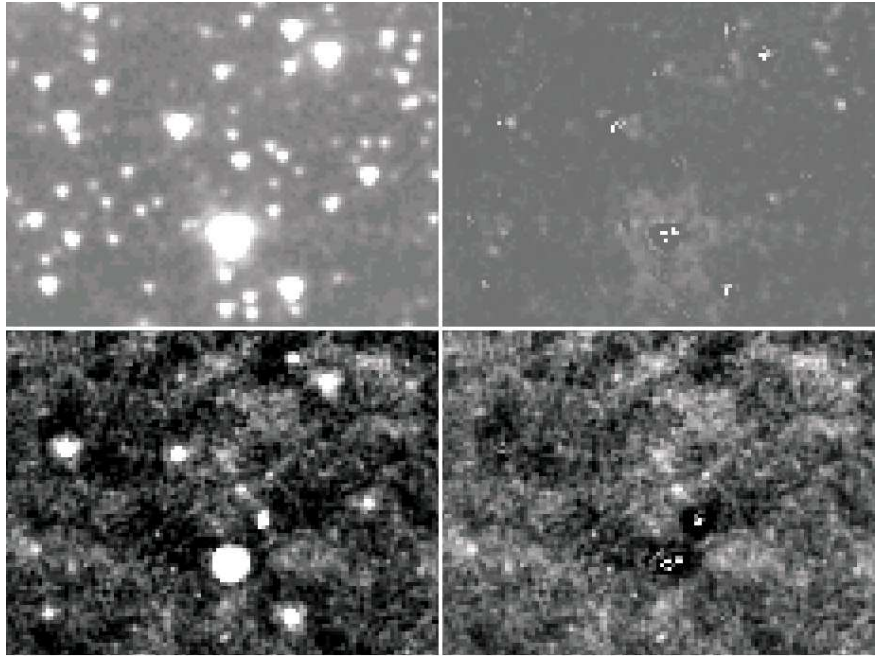}
 \caption{Portion of VMR-D cloud delimited by 
$\alpha$=(08$^h$ 47$^m$ 45$^s$) and $\delta$=(-43\degr 57\arcmin 54\arcsec)  
at the lower left corner, and by $\alpha$=(08$^h$ 47$^m$ 36$^s$) and 
$\delta$=(-43\degr 56\arcmin 41\arcsec) in the upper right. The left panels show 
the mosaics obtained at 3.6~$\mu$m (upper) and 8.0~$\mu$m (lower), respectively. 
The right panels illustrate the residuals remaining after subtraction of the 
point sources that have been modeled with the PRF obtained by using isolated 
unsaturated stars in the same mosaics.} \label{fig_subtracted}
 \end{figure}

\begin{figure}
 \plotone{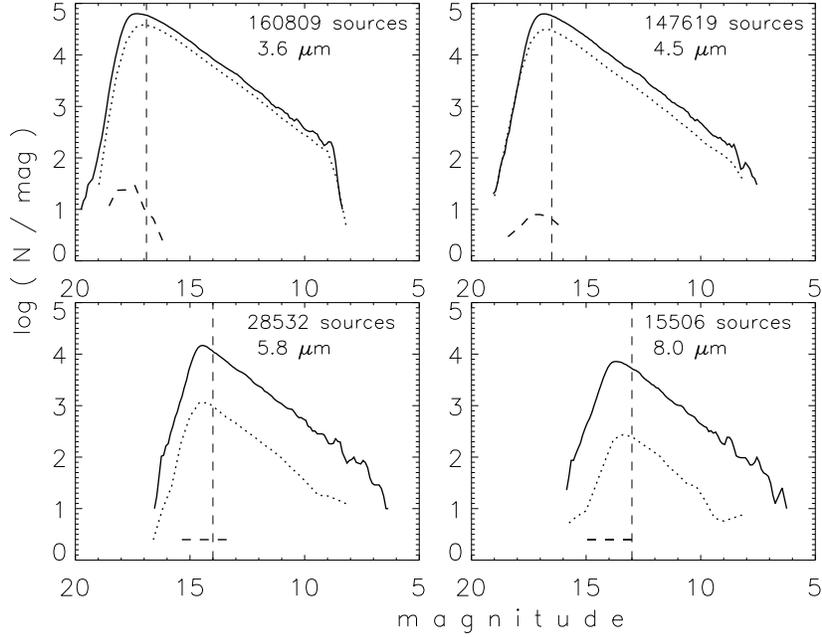}
 \caption{The distribution of the magnitudes in the VMR-D catalog. Each panel 
corresponds to an IRAC band, as indicated. The upper curve (continuous line) 
represents all the sources included in the catalog. The dotted line refers 
to sources observed in both observing epochs (i.e. falling within the green 
dashed line in Figure~\ref{fig_mosaic}). Here the sources included are 
either detected twice or even once but with a detection in an adjacent band. 
The lower dashed curve corresponds to sources considered unreliable because 
they are detected only once, without any detection in adjacent bands. 
From these distibutions the catalog completeness and the reliability are 
estimated (see text). The vertical dashed line is drawn to mark the completeness 
limit. 
 \label{hist_mag}}
\end{figure}

\begin{figure}
 \epsscale{.80}
 \plotone{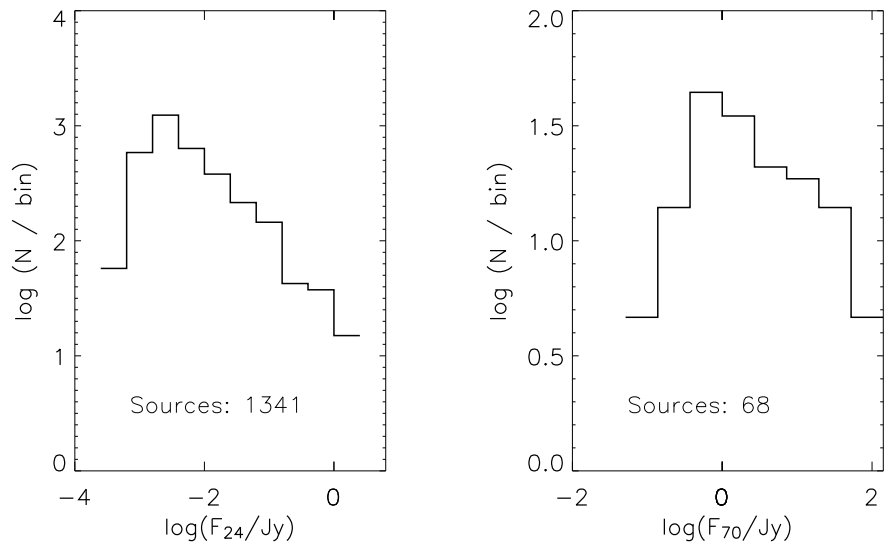}
 \caption{Flux histograms for the VMR-D point sources detected 
 in the MIPS mosaics at 24~$\mu$m (left panel) and 70m~$\mu$m (right panel). 
 In the left panel, 6 saturated sources corresponding to IRS objects have 
 been discarded (see text). \label{hist_MIPS}}
 \end{figure}

\begin{figure}
 \epsscale{1.0}
 \plottwo{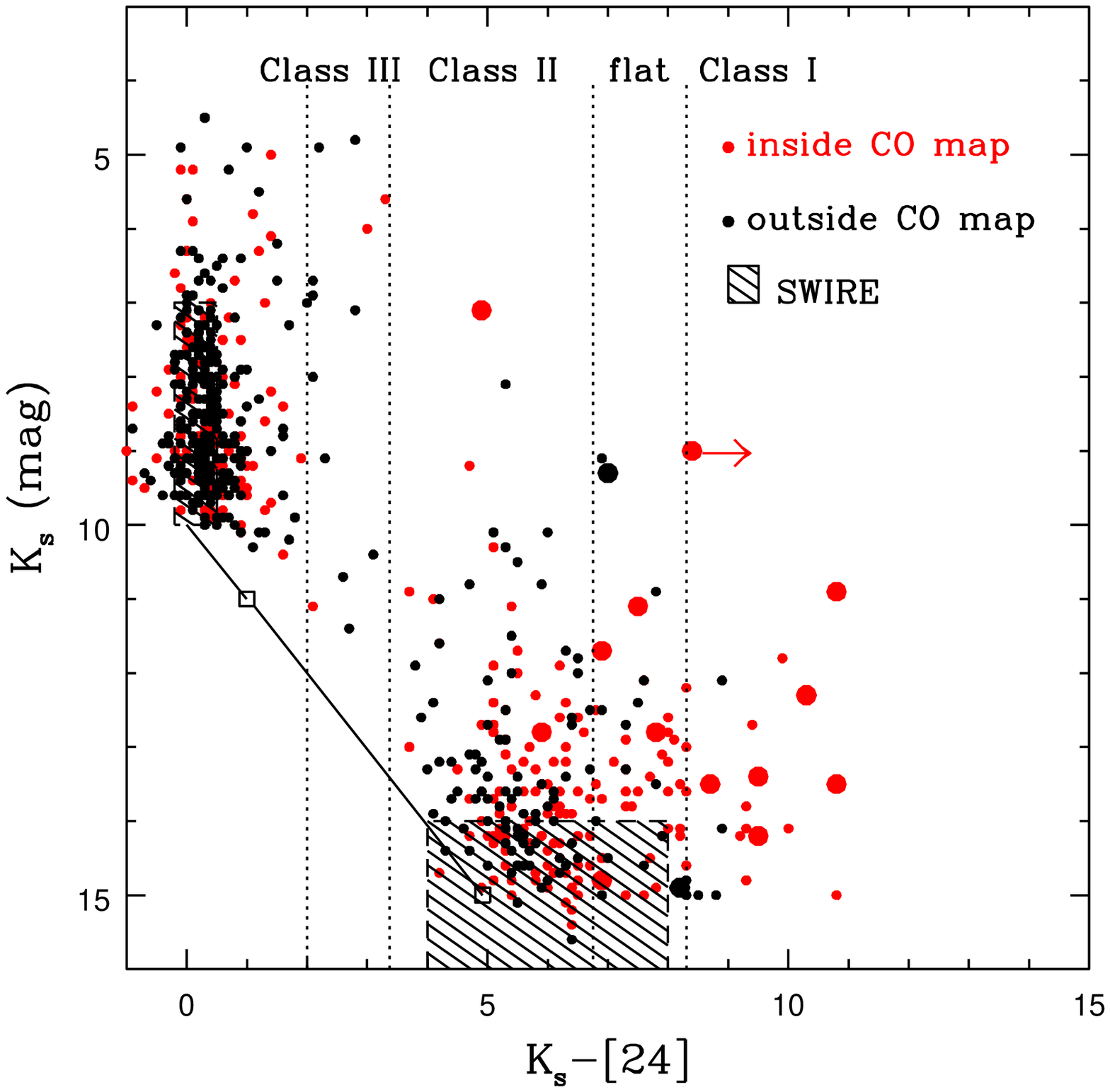}{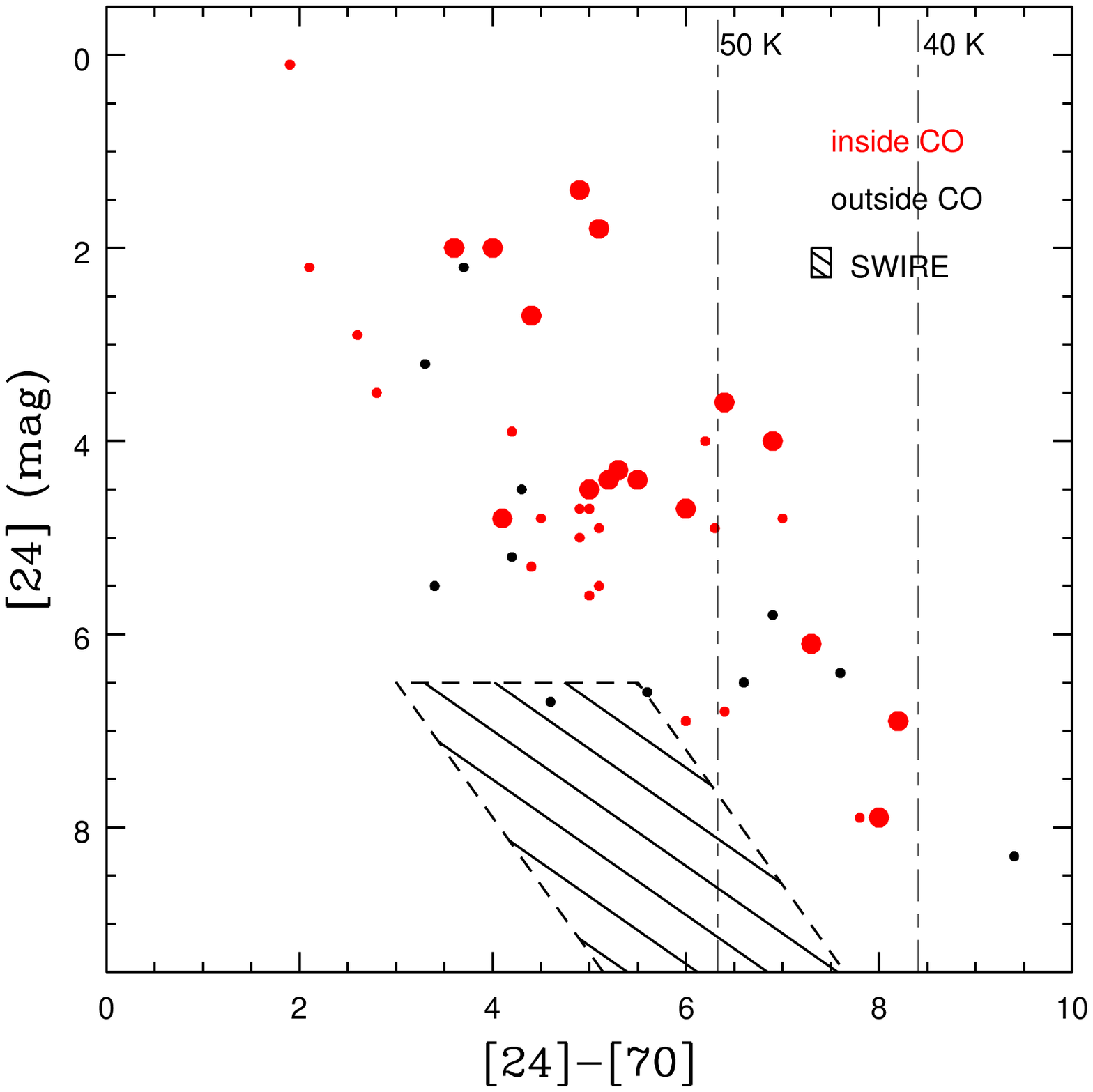}
 \caption{Color-magnitude diagrams for MIPS sources. A comparison with similar diagrams in 
  \citet{Gia07} shows that  by adopting PSF instead of aperture photometry, a systematic 
  decrease ($\lesssim$ 50\%) of the 24~$\mu$m flux and random variations (to a lower extent) 
  of the 70 $\mu$m one has occurred. 
  As a consequence, the number of Class I sources, estimated through 2MASS and 24~$\mu$m 
  photometry (left panel), is now diminished with respect to the early prediction 
  (see also $\S$~\ref{comp_MIPS_IRAC}). Moreover, because of the correspondingly larger 
  [24]-[70] color, an increased number of objects now populates the area corresponding 
  to T $<$ 50K, in the right panel. 
  \label{K_24+24_70}}
 \end{figure}

\begin{figure}
 \epsscale{0.9}
 \plotone{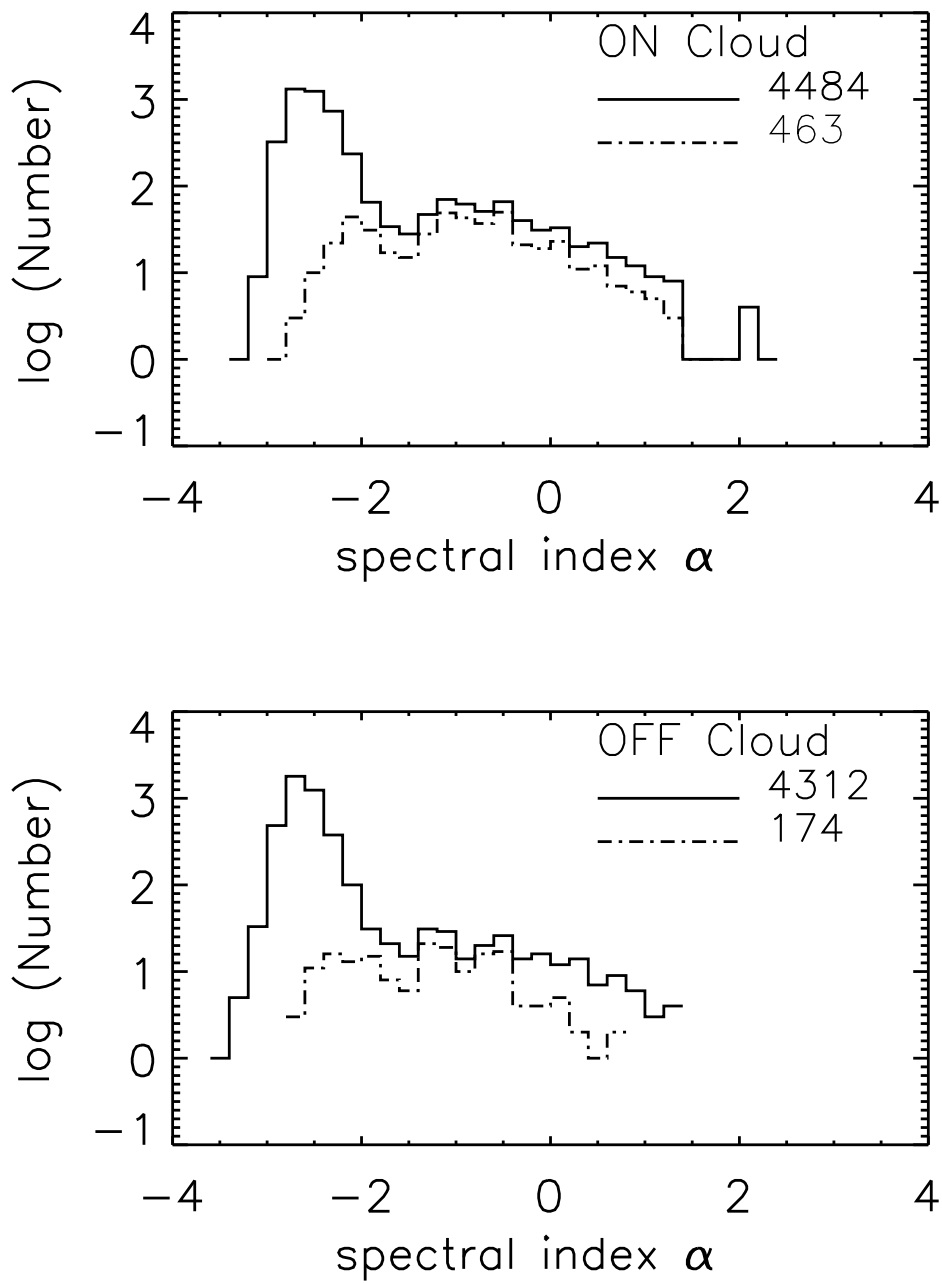}
 \caption{The distribution of the spectral index $\alpha$, as derived by fitting 
 the source fluxes between 2 and 24~$\mu$m (see text). In both panels the continuous 
 line refers to the 8796 ``working sample'' sources (4484 ON-cloud, and 4312 OFF-cloud) 
 with good photometry in the four IRAC bands. 
 After subtraction of both normal stars and extragalactic 
 contamination (see text and Figures~\ref{col_mag_col_ON}, \ref{col_mag_col_OFF} and 
 \ref{col_mag_col_ONexclu}) we remain with 637 sources, that are located either 
 ON-cloud (dashed line, 463 cases) or OFF-cloud (dashed line, 174 cases). }
 \label{hist_alpha} 
 \end{figure}

\begin{figure}
 \epsscale{0.9}
 \plotone{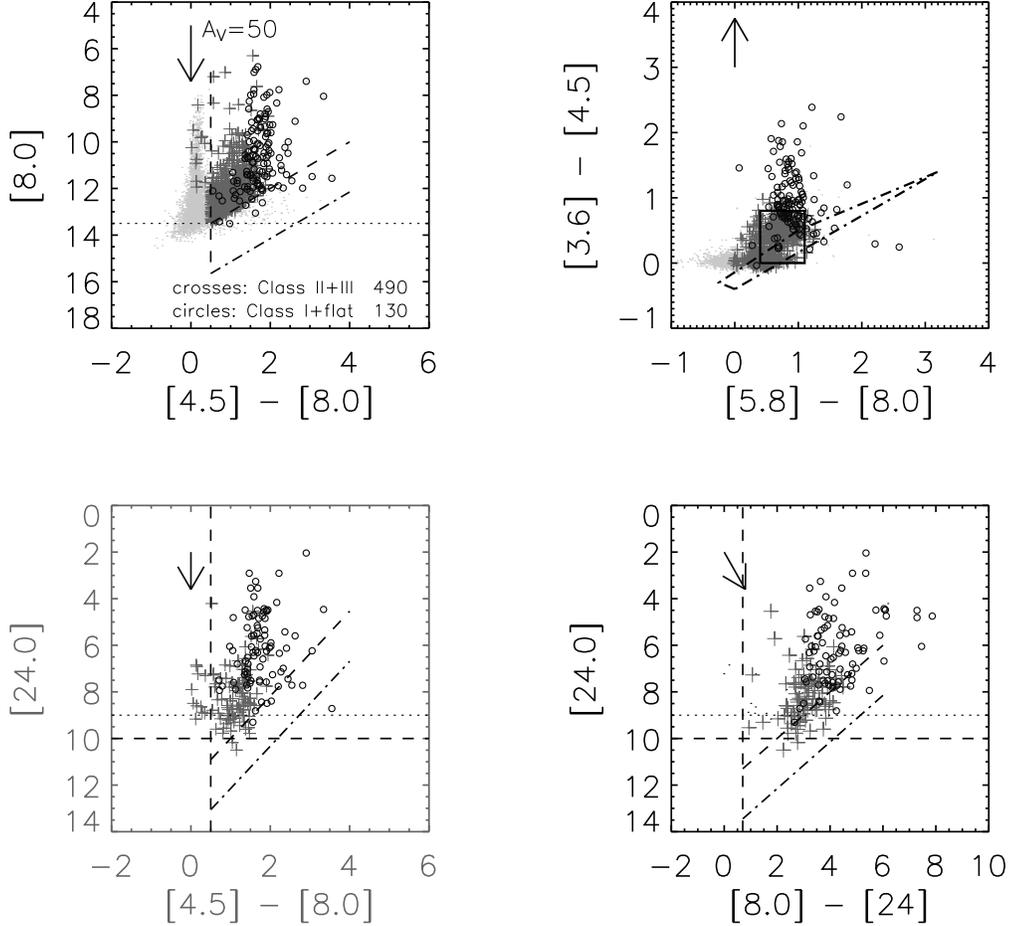}
 \caption{ Color-magnitude and color-color diagrams for the VMR-D point 
 sources detected in the four IRAC bands and located ON-cloud (see \S~\ref{obse}). 
 Sources selected applying the \citet{Har06,Har07} criteria are plotted as circles 
 (spectral index $\alpha > -0.3$) and grey crosses ($\alpha < -0.3$). In the upper 
 panels grey dots represent the 8796 ``working sample'' sources.  
 The reddening effect, corresponding to a visual extinction of A$_V$=50 mag, 
 is shown by the arrow \citep{Ind05}. 
 In the color-magnitude plots the dashed lines correspond to statistical criteria 
 derived by \citet{Har06,Har07} for separating YSO candidates from normal stars 
 (vertical lines) and extragalactic contamination (oblique lines) in the Serpens dark 
 cloud. The dash-dotted line represents the displacement of the dashed line 
 after scaling to the VMR-D distance (see text). 
 The dotted lines show the completeness limit 
 and the horizontal dashed line shows the 24 $\mu$m magnitude beyond which a 
 source is considered extragalactic. 
 In the two-color upper right panel the small square shows the approximate 
 domain for the Class II sources according to \citet{All04}. 
 The dot-dashed line encloses the region corresponding to the AGB colors, 
      but note that a source can be considered as a candidate AGB only when two 
      further constraints are satisfied in the  [3.6]-[4.5] vs [3.6]-[8.0]
 and [4.5]-[5.8] vs [5.8]-[8.0] diagrams, respectively \citep{Mar08}. 
 \label{col_mag_col_ON} } 

 \end{figure}

\begin{figure}
 \epsscale{0.9}
 \plotone{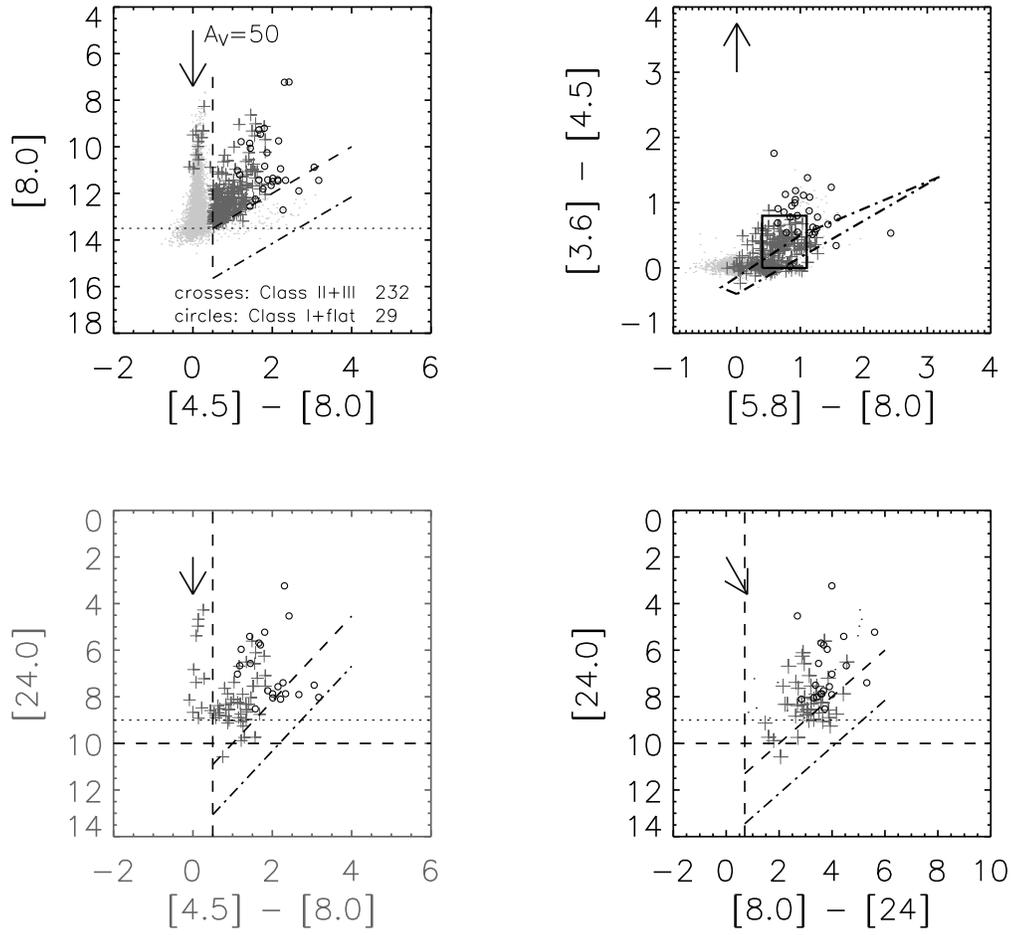}
 \caption{The same as in Fig.~\ref{col_mag_col_ON} but for sources located 
 OFF-cloud. \label{col_mag_col_OFF}}  
 \end{figure}

\begin{figure}
 \epsscale{0.9}
 \plotone{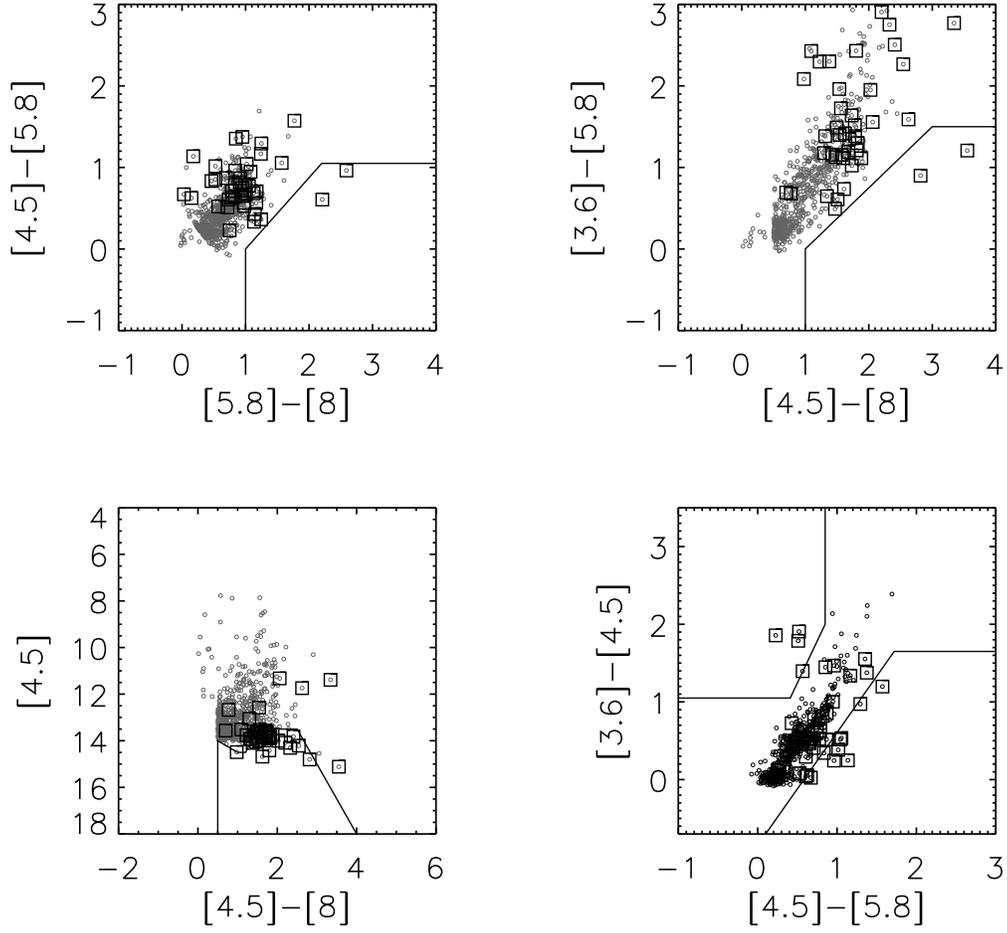}
 \caption{ 
 Diagrams used to isolate the contaminants among the sources 
 of the ``working sample'' selected with the \citet{Har06,Har07} criteria. 
 For simplicity only ON-cloud sources are displayed. In each diagram, solid 
 lines delimit the loci where possible contaminants are located \citep[see][]{Gut09}.
 The grey circles represent the sources shown in Fig.~\ref{col_mag_col_ON} 
 as crosses and circles. The upper panels are used to isolate 
 unresolved starforming galaxies falling under the continuous line. The lower 
 left panel isolates broad-line AGN and the lower right is used to 
 identify likely shock emission (upper left region) and possible cases of 
 PAH contamination (lower right). 
 Objects falling in at least one of the regions delimited by the continuous 
 lines are identified as likely contaminants and are reported as open squares. 
 \label{col_mag_col_ONexclu}}
 \end{figure}

We applied the $w(\theta)$ statistics to our candidate YSOs selected among the 
VMR-D working sample sources (see \S~\ref{YSO_class}); that is, those sources 
satisfying the criteria for minimizing the presence of galactic and extragalactic 
contaminants illustrated in \S~\ref{source_selection} and Figures~\ref{col_mag_col_ON}, 
\ref{col_mag_col_OFF}, and \ref{col_mag_col_ONexclu}. 
The spatial distribution of these sources is presented in Figure~\ref{vela_YSO+CO} 
where the objects of Class I/flat (red circles) and Class II/III (green crosses) are 
shown superposed to the 8.0 $\mu$m IRAC mosaic. 

\begin{figure}
 \epsscale{1.}
 \plottwo{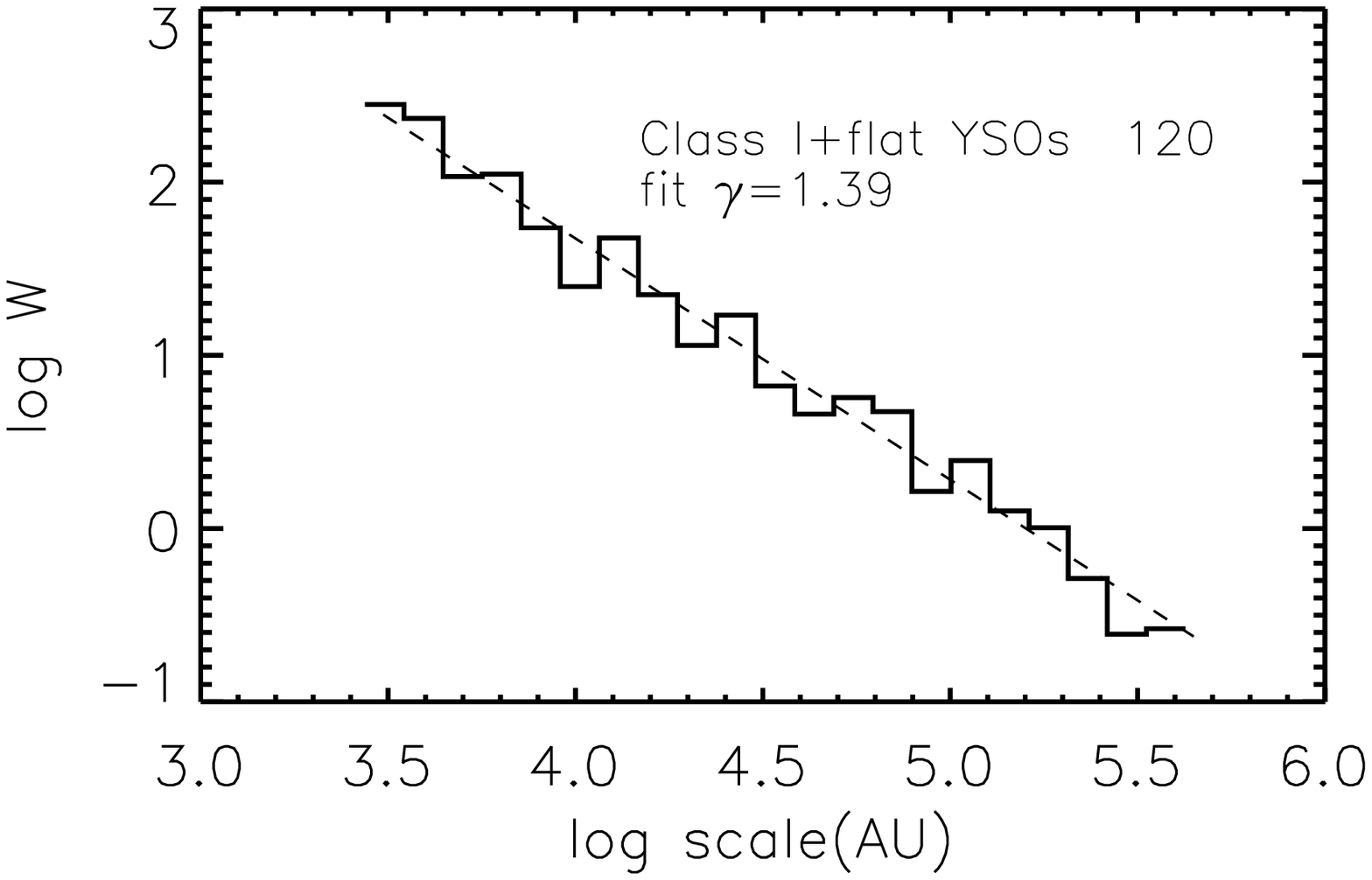}{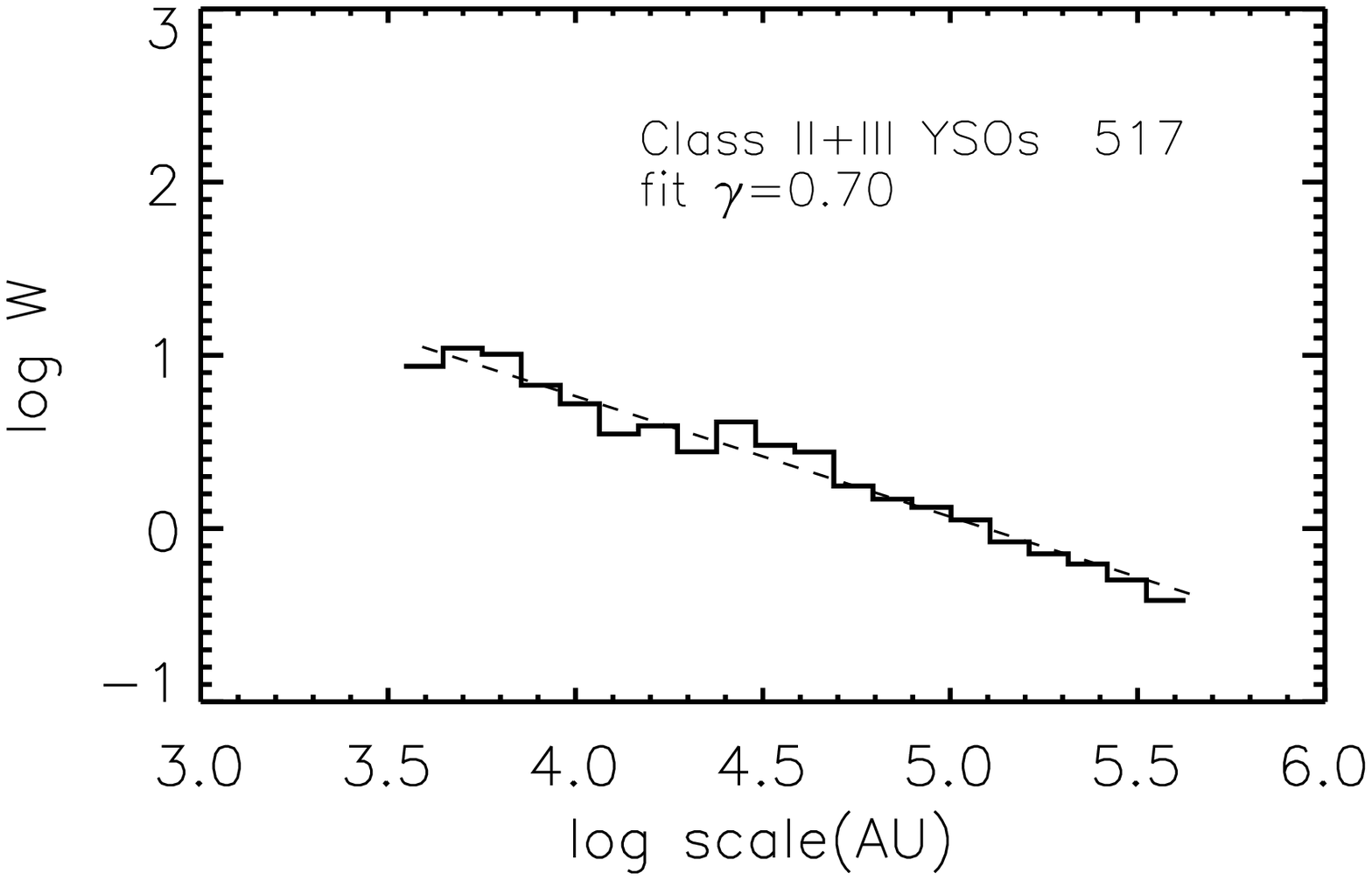}
 \caption{The two-point correlation function for all the candidate YSOs 
  in VMR-D cloud. Left panel: sources classified as Class I and flat-spectrum. 
  Right panel: Class II and III sources. The number of objects is reported in each 
  panel with the slope $\gamma$ of the linear fit. The slope uncertainty, evaluated 
  by bootstrap resampling, is $\pm0.10$ and $\pm0.08$ in the 
  left and right panel, respectively.  }
 \label{acf} 
 \end{figure}

We find substantial differences between clustering in early and late Classes as 
is shown in Figure~\ref{acf} where we present the behaviour of the $w(\theta)$ function 
for Class I and flat-spectrum (left panel), and Class II and III sources (right 
panel), respectively. 
It is apparent that in both cases the correlation function behaves as a single power 
law, at least in the investigated range of scales, but with noticeably different 
slopes denoting a higher degree of clustering of the younger classes with respect 
to the later evolutionary stages. This can be also qualitatively seen by eye from 
the spatial distributions in Figure~\ref{vela_YSO+CO}. 

Note, however, that in the range of the investigated spatial scales, there is 
no break in the power law slope within each region; this is significant 
even though a direct comparison 
with the cases discussed in \citet{Sim97} for optically selected PMS stars is not 
straightforward because here we deal with IRAC selected YSOs instead of optically 
selected PMS stars. A more appropriate comparison can be done with \citet{Har07} who 
show the behavior of the same $w(\theta)$ function computed for the distribution of 
the YSOs in Serpens (their Figure 10).  The slopes they found for both Class I/flat 
and Class II/III sources are remarkably similar to those shown in Figure~\ref{acf} 
for VMR-D sources. A possible difference may be due to a trend in Serpens 
toward smaller scales due to a flattening of the function at scales smaller  
than $\approx 5\times10^3$ AU. 

A plausible justification for the different slopes of early and late YSOs is simply 
that it is due to the different ages of the two populations. It is generally accepted 
that stars are essentially formed in clusters, so that it is naturally expected that 
the older objects have moved from their birthplaces, producing object dispersal we 
measure in the shallower slope of the correlation function for Class II/III sources. 

\begin{figure} 
 \epsscale{0.9}
 \plottwo{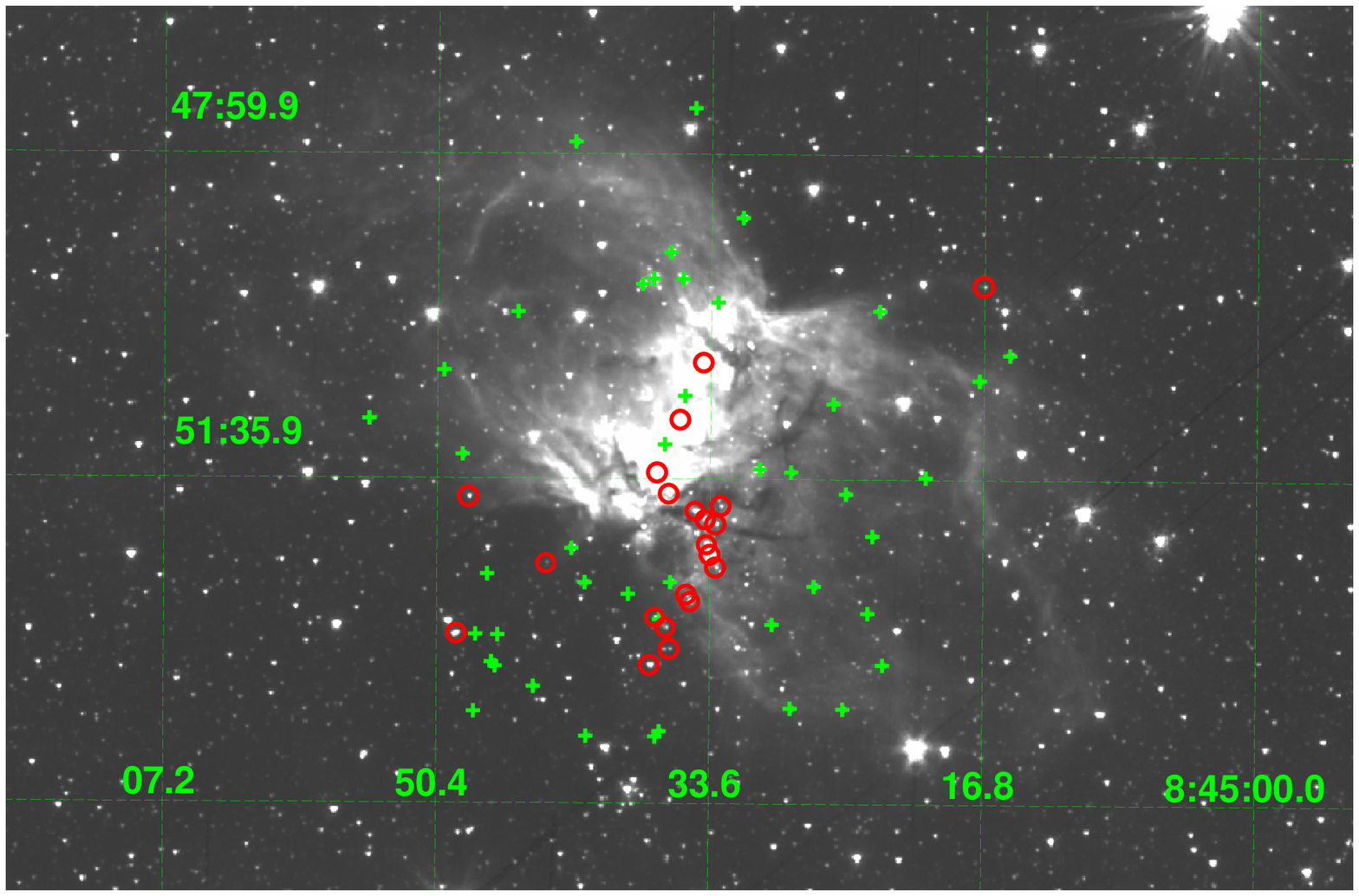}{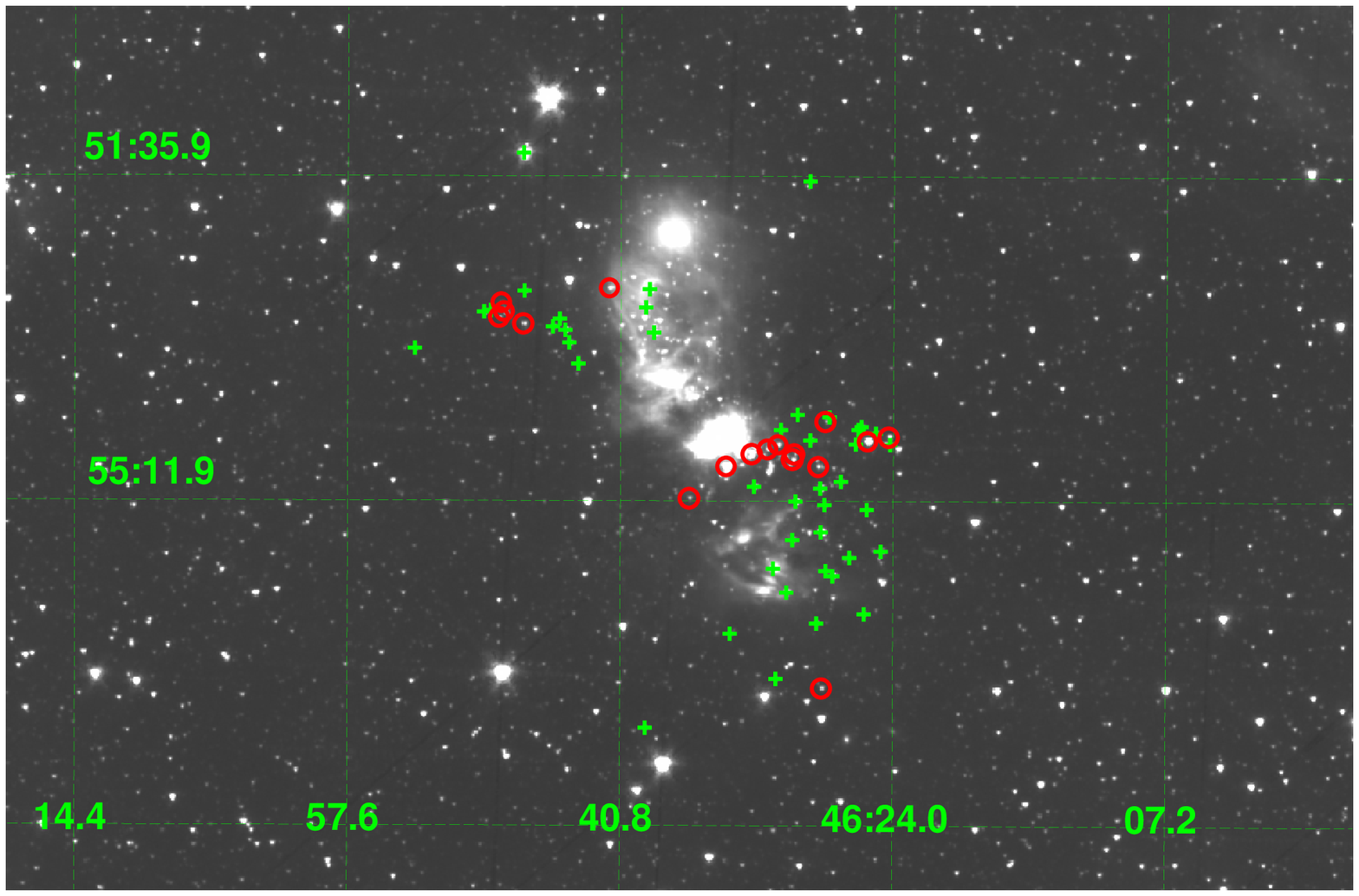}
 \caption{Class I and flat-spectrum (red circles) and Class II and III (green crosses) 
 IRAC sources found in a radius of 4$^\prime$~around IRS16 (left panel) and IRS17 
(right panel), respectively. Sources shown here possess at least three IRAC flux 
 densities of good quality (see text). \label{thumb_16_17}}
 \label{thumb} 
 \end{figure}

\begin{figure}
 \epsscale{0.9}
 \plottwo{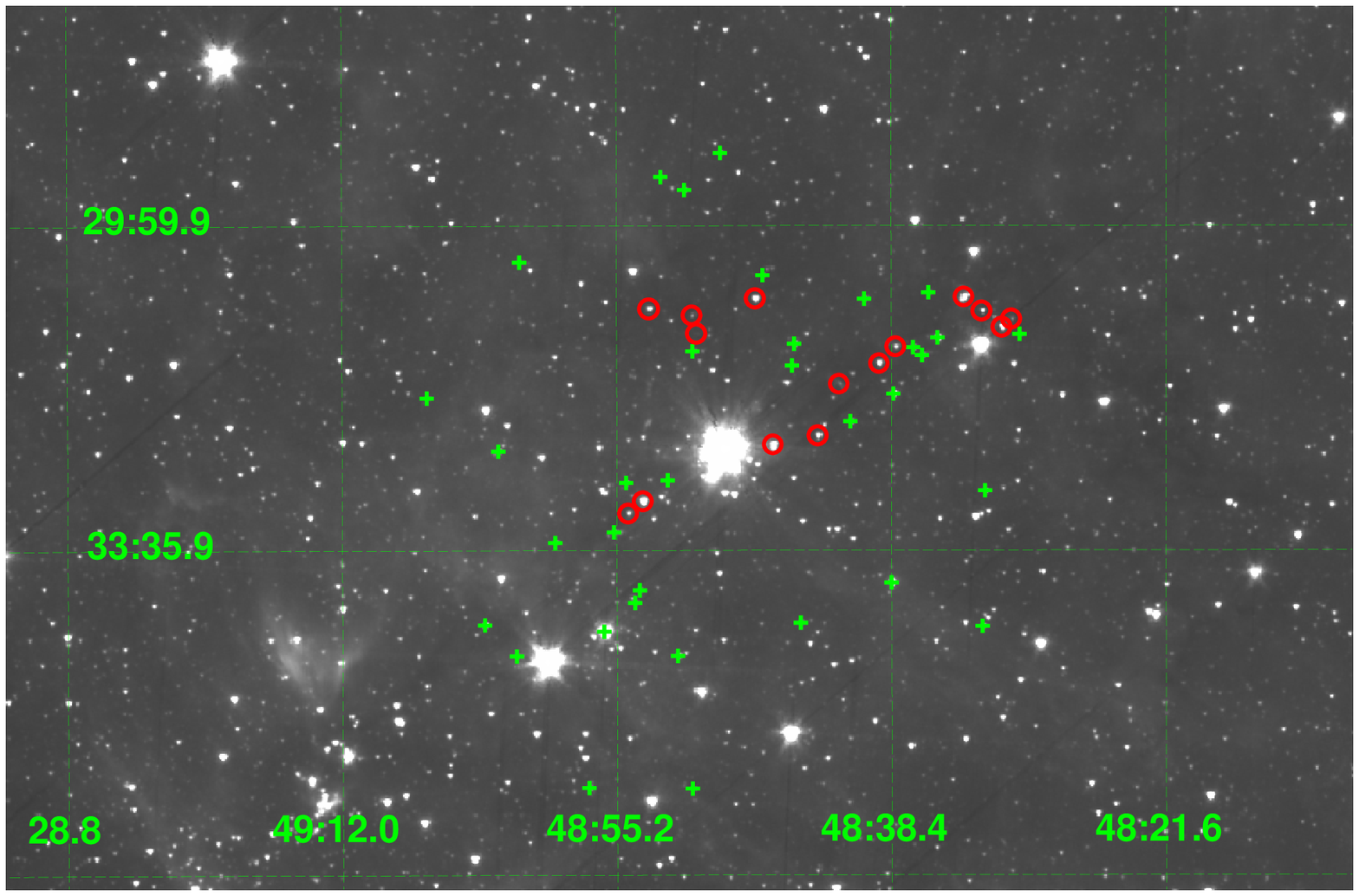}{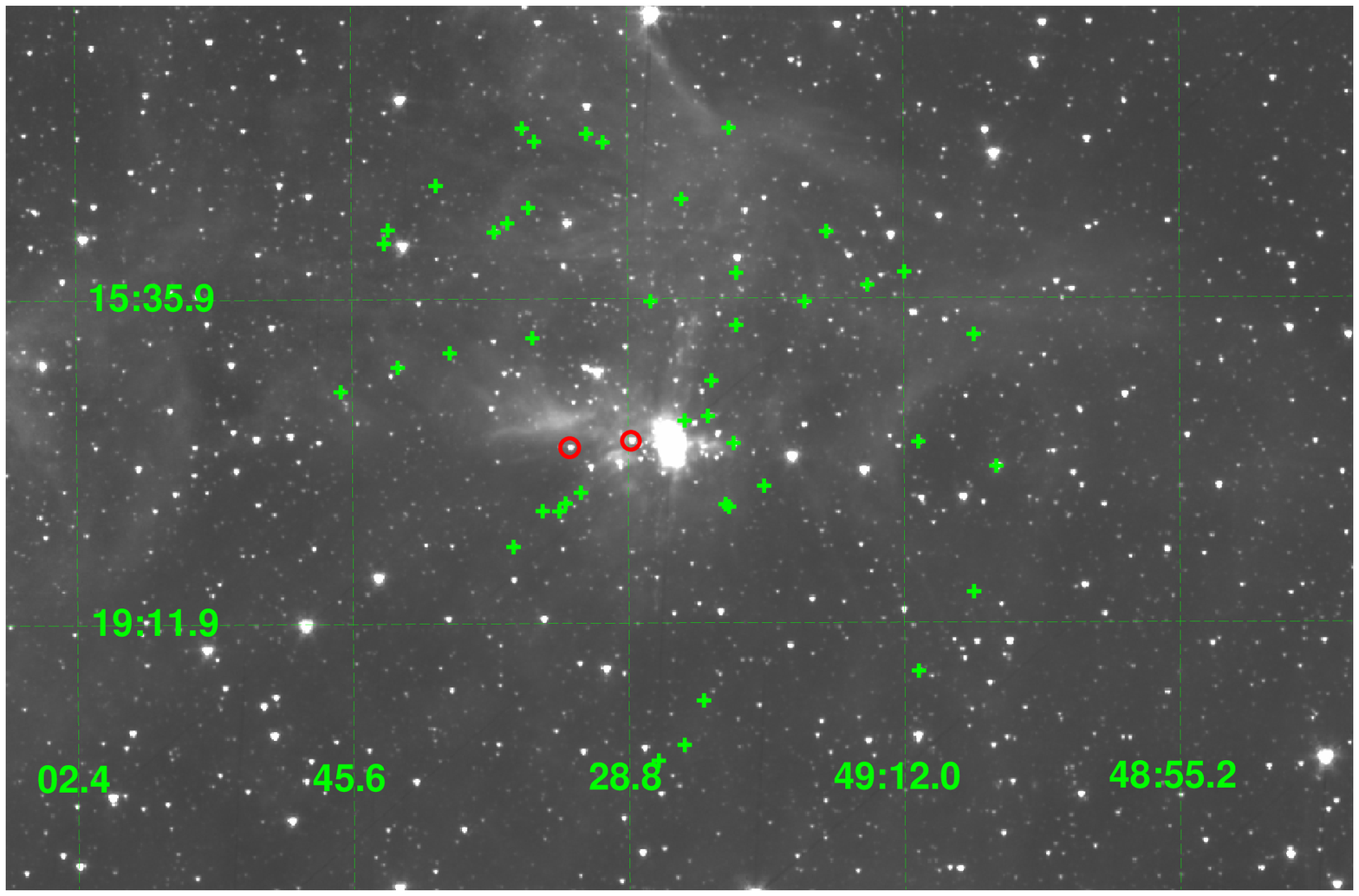}
 \caption{As in Figure~\ref{thumb} but for the regions around IRS19 (left panel) and IRS20 
(right panel), respectively. \label{thumb_19_20}}
 \end{figure}

Finally in Figures~\ref{thumb_16_17} and \ref{thumb_19_20} we show more details of the 
four regions around IRS16, 17, 19, and 20, including the spatial distributions of the 
Class I/flat (red circles) and Class II/III (green crosses) located within 4$^\prime$. 
To produce these figures, we consider an enlarged sample containing all the 
catalog sources located around the IRS objects and with good photometry in at least 
two adjacent IRAC bands. This allows us to compensate for the relative scarcity in 
these small areas of candidate YSOs selected from our working sample and then to 
better highlight the trend in the spatial distribution of the different 
classes. 
The sources reported here have been selected from this enlarged sample by applying 
the same criteria discussed in \S~\ref{source_selection} and illustrated in  
Figure~\ref{col_mag_col_ON}, \ref{col_mag_col_OFF}, and \ref{col_mag_col_ONexclu}. 
In doing this we discarded the sources with upper limits and considered only 
those cases for which at least a bad photometry was obtained in the two 
complementary IRAC bands. 

In all cases we see that around these four IRS objects the Class I/flat sources 
(red circles) are not uniformly distributed.  They appear more abundant around 
IRS16 and IRS17; Class II/III objects (green crosses) are instead more numerous and 
more simmetrically distributed in IRS16, 19, and 20. 
An exception is IRS17 where the Class II/III objects follow the same 
elongated distribution of the Class I/flat sources. We also note that this 
configuration in IRS17 is suggestive of the presence of two separated clusters as 
has also been suggested by \citet{Mas06}.  
Speculating on this case, we suggest that a dust disk, whose plane is perpendicular to 
the observed YSOs distribution, may be responsible for this appearance; this is suggested 
also by the central dark lane visible in all the IRAC bands. It is however clear that a 
conclusion cannot be given here and further investigations are needed.

\section{Conclusions}  \label{concl}
This paper presents the analysis of the {\em Spitzer}-IRAC mosaics obtained 
in the region of the Vela Molecular Ridge, Cloud D. In a $\approx$~1.2 square 
degrees area, corresponding to the overlap of the IRAC mosaics in the four bands, 
we obtained the photometry that has been used to compile a four-band point source 
catalog. This has been supplemented by relevant information at longer wavelengths 
from revised photometry of the VMR-D MIPS mosaics and from previously 
published millimeter line and continuum observations of the same region. 
This catalog has been used to outline the general characteristics of the YSOs 
in this SF region. The main results are: 
\begin{enumerate} 
\item The VMR-D IRAC point source catalog has been compiled with 170,299 entries, 
      8796 of which, constituting our working sample, have good quality fluxes in 
      all four IRAC bands. 
      Complementary information has been also added using a revised, PSF-based, 
      MIPS 24~$\mu$m and 70~$\mu$m photometry and previous millimeter observations 
      obtained in both CO lines and dust continuum. 
\item A revised catalog containing the point sources extracted from the VMR-D MIPS 
      24~$\mu$m and 70~$\mu$m mosaics has been also obtained following the same photometric 
      procedure adopted for the IRAC bands. 
\item The working sample has been analyzed according to the spectral slope 
      that was obtained by also considering both the K$_s$-2MASS and the MIPS 24~$\mu$m
      fluxes when available. 
\item We selected the candidate YSOs by means 
      of their position in  color-color and color-magnitude diagrams 
      involving IRAC and MIPS photometry; the likely contaminants due to 
      extragalactic and galactic objects have been also estimated and excluded 
      from the final list of YSOs constituted by 637 objects. 
\item The candidate YSOs have been used to evaluate the surface density
      (3.3--5.5 pc$^{-2}$), the star formation rate 
      (2.5--4.1 M$_\sun$Myr$^{-1}$pc$^{-2}$) 
      and efficiency ($\approx$~0.02--0.03), 
      and the depletion time (40--70 Myr) of the VMR-D cloud. 
\item The global clustering properties of the YSOs, studied by means of the two point 
      correlation function, have been found to be very similar to those derived by 
      \citet{Har07} for the Serpens SF region; it clearly shows that Class I/flat 
      sources are more clustered than Class II/III sources. 
\item The IRAC catalog sources around IRS16 and IRS17 show a larger fraction of   
      Class I/flat SEDs with respect to IRS19 and IRS20 where the 
      Class II/III objects are relatively more numerous.  
      In the IRS17 region there is a tendency for the Class II/III to trace the same 
      elongated distribution of the Class I/flat sources. 
\end{enumerate}



\acknowledgments

We thank Joao Lin Yun, Luca Olmi and Berlinda Maiolo for many helpful discussions. 
DE has been partly supported by the European Commission FP6 Marie Curie Research 
Training Network “CONSTELLATION” (MRTN-CT-2006-035890). FS aknowledges partial 
support by Italian Space Agency (ASI). 
This work is based on observations made with the Spitzer Space Telescope, 
which is operated by the Jet Propulsion Laboratory, California Institute of Technology 
under a contract with NASA. Support for this work was provided by NASA through an 
award issued by JPL/Caltech.

\clearpage

 \begin{deluxetable}{lcccccc}
 \tablecaption{Overview of the limiting magnitudes and completeness\label{tab_lim_fluxes}}
 \tablewidth{0pt}
 \tablehead{
            & \multicolumn{4}{c}{$IRAC$} & \multicolumn{2}{c}{$MIPS$}  \\
 \colhead{} & \colhead{3.6} & \colhead{4.5} & \colhead{5.8} & \colhead{8.0} 
            & \colhead{24}  & \colhead{70} 
 }
 \startdata 
 F$_{\rm lim}$\tablenotemark{a} & $1.1\times10^{-5}$  & $1.0\times10^{-5}$  & $3.5\times10^{-5}$  & $4.9\times10^{-5}$  & $6.3\times10^{-4}$ &  $1.0\times10^{-1}$  \\
 m$_{\rm lim}$                  &  18.5   &  18.1   & 16.3    &  15.3   & 10.1   &  2.2     \\
 F$_{\rm comp}$\tablenotemark{a}& $4.9\times10^{-5}$  & $4.5\times10^{-5}$  & $2.9\times10^{-4}$  & $4.0\times10^{-4}$  & $2.0\times10^{-3}$ &  $5.0\times10^{-1}$    \\
 m$_{\rm comp}$                 &  16.9   & 16.5    & 14.0    &  13.0   & 8.9    &  1.0  
 \enddata
 \tablenotetext{a}{All fluxes are given in Jy}
 \end{deluxetable}

\begin{deluxetable}{lcccccccc}
\tabletypesize{\footnotesize}   
\tablecaption{VMR-D MIPS catalog: typical entries.\tablenotemark{a} \label{tab_MIPS_cat}}
\tablewidth{0pt}
\tablehead{\colhead{MIPS-ID} & \colhead{RA(2000)} & \colhead{Dec(2000)}  & \colhead{Glon}  & \colhead{Glat}  & \colhead{F24} & \colhead{E24} & \colhead{F70} \\
                             & \colhead{E70}      & \colhead{$\Delta\theta$\tablenotemark{b}} &  \colhead{IRAC\tablenotemark{c}} & \colhead{$^{12}$CO(1-0)\tablenotemark{d}} & \colhead{$^{13}$CO(2-1)\tablenotemark{d}} & \colhead{Dust\tablenotemark{d}}   \\
          }
\startdata      
            1                & 130.79321          & -44.01385            &   263.47831     & -0.97396        &  8.420e-03    & 1.812e-03     &  0.000e+00 \\ 
                             & 0.000e+00          &   0.0                &      N          &        N        &   N           &     -     \\
            2                & 130.81630          & -43.97870            &   263.46093     & -0.93919        &  1.535e-03    & 1.135e-04     &  0.000e+00 \\ 
                             & 0.000e+00          &   0.0                &     N           &        N        &   N           &     -     \\
\nodata & \nodata   & \nodata    & \nodata   & \nodata  &  \nodata   & \nodata    & \nodata     \\ 
         135                 & 131.38707          & -43.71376            &  263.50867      & -0.45199        &  4.314e-03    & 1.037e-04     & 0.000e+00  \\
                             & 0.000e+00          &   0.0                &     I           &      N          &   N           &      -    \\
         136                 & 131.38847          & -43.83051            &  263.60060      & -0.52397        & 1.934e+00     & 2.598e-01     &  2.001e+01 \\ 
                             & 8.478e-01          &   2.1                &    I            &  C              &  C            & MMS1      \\  
\nodata & \nodata   & \nodata    & \nodata   & \nodata  &  \nodata   & \nodata    & \nodata     \\ 
\enddata
\tablenotetext{a}{Coordinates are given in degrees; fluxes (F) and their uncertainties (E) are given 
                  in Jy.}
\tablenotetext{b}{Angular distance in arcsec between 24~$\mu$m and 70~$\mu$m centroids.} 
\tablenotetext{c}{I and N are used to flag a source as falling inside or outside the IRAC observed region.}
\tablenotetext{d}{Symbols are explained in  \S~\ref{thePSC} (Columns 26-28).}

\end{deluxetable}

\begin{deluxetable}{lcccc}
 \tablecaption{MIPS source statistics \label{tab_stat_MIPS}}
 \tablewidth{0pt}
\tablehead{\colhead{CO association} &  \colhead{24~$\mu$m only} & \colhead{70~$\mu$m only} & \colhead{24~\&~70~$\mu$m} & \colhead{Total}  }
\startdata 	
Inside contours      &  604   &  7  &  40\tablenotemark{a}  &  651 \\
Outside contours     &  131   &  1  &   2                   &  134 \\
Outside IRAC catalog &  560   &  8  &  10\tablenotemark{b}  &  578 \\
\tableline
Total                & 1295   & 16\tablenotemark{c}  &  52                   & 1363  \\
\enddata
\tablenotetext{a}{Five sources are saturated at 24 $\mu$m corresponding to 
 IRS16, 17, 19, 20, 21; five other sources have a duplicate 70 $\mu$m association. }
\tablenotetext{b}{One source is saturated at 24 $\mu$m, corresponding to IRS18. }
\tablenotetext{c}{Four sources are outside the 24~$\mu$m mosaic; visual inspection 
       of the remaining shows that nine are diffuse while three appear more compact 
       and are reported in Table~\ref{tab_MIPS_remark} as remarkable objects.}
\end{deluxetable}

\begin{deluxetable}{lccccccccc}
\tablewidth{0pt}
\tablecaption{Classification based on MIPS and IRAC photometry \label{tab_class_MIPS-IRAC}}
\tablehead{      &   \multicolumn{2}{c}{ MIPS-based }  &  & \multicolumn{2}{c}{ IRAC-based } \\
\colhead{Class}  &  \multicolumn{1}{c}{ ON-cloud }  & \multicolumn{1}{c}{ OFF-cloud} &  & \multicolumn{1}{c}{ ON-cloud }  & \multicolumn{1}{c}{ OFF-cloud} \\ 
                              &  \colhead{N (\%)\tablenotemark{a}} & \colhead{N (\%)\tablenotemark{a}} &  & \colhead{N (\%)} & \colhead{N (\%)}  }
\startdata
Class I       &   16 (10.1)   &     4 (3.4)   &  &  44 (9.5)  &   5 (2.9)    \\ 
Flat spectrum &   41 (25.9)   &    20 (16.9)  &  &  61 (13.2) &  10 (5.7)    \\ 
Class II      &   98 (62.0)   &    84 (71.2)  &  & 230 (49.7) &  93 (53.4)   \\ 
Class III     &    3 (1.9)    &    10 (8.5)   &  & 128 (27.6) &  66 (37.9)   \\ 
\enddata
\tablenotetext{a}{These MIPS-based percentages can be compared with those in Tab.~5 of \citet{Gia07} 
after renormalization of their values to the YSOs only.}  
\end{deluxetable}

 \begin{deluxetable}{lccccccccccccccccccccccccccc}
 \tabletypesize{\scriptsize}
 \tablecaption{VMR-D IRAC Point Source Catalog: typical entries \tablenotemark{a}. \label{tab_IRAC_PSC}}
 \tablewidth{0pt}
 \tablehead{ \colhead{ID}  & \colhead{RA(2000)} & \colhead{Dec(2000)} &  \colhead{Glon}  & \colhead{Glat}      & \colhead{F3.6} &   \colhead{E3.6} &  \colhead{Q3.6}  \\  
                           & \colhead{F4.5} & \colhead{E4.5}  &  \colhead{Q4.5}    &  \colhead{F5.8}           & \colhead{E5.8} & \colhead{Q5.8} & \colhead{F8.0}  \\  
			   & \colhead{E8.0} & \colhead{Q8.0}  &  \colhead{MIPS ID} & \colhead{$\Delta\theta$\tablenotemark{b}}  & \colhead{F24} & \colhead{E24} & \colhead{Q24} \\
			   & \colhead{F70}  & \colhead{E70}   & \colhead{Q70}      & \colhead{$^{12}$CO(1-0)\tablenotemark{c}}  & \colhead{$^{13}$CO(2-1)\tablenotemark{c}}  &  \colhead{Dust\tablenotemark{c}} \\
 }
 \startdata  
      1  & 131.031909  & -43.736558  & 263.366966 & -0.667202   & 1.89903e-05 & 2.54204e-06 & L1 \\
         & 4.15450e-06 & 0.00000e+00 & u          & 2.11600e-05 & 0.00000e+00 & u           & 2.87050e-05  \\ 
	 & 0.00000e+00 & u           &  -         & 0.0         & 3.084e-04   & 0.000e+00   & u   \\
	 & 5.220e-04   & 0.000e+00   & u          & N           & N           & -  \\
\nodata &  \nodata   &  \nodata     & \nodata     &  \nodata   &  \nodata     & \nodata     &  \nodata   \\ 
    393  & 131.096329  & -43.775259  & 263.426188 & -0.654711   & 1.32000e-02 & 7.87000e-04 & b  \\
         & 8.33412e-03 & 2.59158e-04 & H1         & 5.53000e-03 & 3.59000e-04 & b           &  3.56018e-03 \\
	 & 7.23856e-05 & H1          & 37         & 1.7         & 2.543e-02 & 3.492e-03     &  G  \\
	 & 3.907e-04   & 0.000e+00   & u          & N           & N         & -  \\ 
 \nodata &  \nodata   &  \nodata     & \nodata     &  \nodata   &  \nodata     & \nodata     &  \nodata   \\ 
   2548  & 131.183726  & -43.701613  & 263.407676 & -0.559483   & 1.42000e-03  & 8.73000e-05 & b   \\ 
         & 9.11296e-04 & 3.42903e-05 & H2         & 6.48618e-04 & 5.45410e-05  & H1          &  3.10650e-04 \\ 
	 & 1.79767e-05 & L2          & 57         & 5.1         & 4.497e-02    & 1.205e-03   &  G  \\
	 & 1.165e-01   & 1.191e-02   & G          & N           & N            & - \\
 \nodata &  \nodata   &  \nodata     & \nodata    &  \nodata   &  \nodata     & \nodata     &  \nodata   \\ 
  17590  & 131.388578  & -43.868336  & 263.630227 & -0.547490   & 2.54057e-03  & 6.21729e-05 & H2  \\
         & 3.98529e-03 & 1.06327e-04 & H2         & 3.52000e-03 & 1.78000e-04  & b          &  1.29000e-03 \\
	 & 3.38000e-04 & b           & 139        & 6.0         & 3.754e-01    & 4.395e-02  &  G   \\
	 & 8.675e-02   & 0.000e+00   & u          & C           & C            & MMS2  \\ 
 \nodata &  \nodata   &  \nodata     & \nodata     &  \nodata   &  \nodata     & \nodata     &  \nodata   \\ 
 170299  & 132.967055  & -43.592335  & 264.135486 & 0.511679    & 5.86857e-05  & 5.13993e-06 & H1  \\
         & 5.54702e-05 & 6.68016e-06 & H1         & 2.14100e-05 & 0.00000e+00  & u           & 3.13800e-05 \\
	 & 0.00000e+00 & u           & -          & 0.0         & 2.574e-04    & 0.000e+00   & u   \\
	 & NaN         & 0.000e+00   & u          & N           & N            & -   \\
 \enddata
\tablenotetext{a}{Coordinates are given in degrees; fluxes (F) and their uncertainties (E) are given 
                  in Jy; quality flags (Q) are described in \S~\ref{thePSC}.}
\tablenotetext{b}{Angular distance in arcsec between the IRAC and MIPS centroids.}
\tablenotetext{c}{Symbols are explained in \S~\ref{thePSC}, when Columns 26-28 are described. } 
\end{deluxetable}

\clearpage

 \begin{deluxetable}{lr}
 \tablecaption{ VMR-D cloud IRAC catalog statistics (global) \label{tab_stat_global}}
 \tablewidth{0pt}
 \tablehead{ \colhead{With good photometry} & \colhead{Number of sources} }
 \startdata 
 In only one IRAC band       &        53461    \\
 In only two IRAC bands      &        97114    \\
 In only three IRAC bands    &        10851    \\
 In all four  IRAC bands\tablenotemark{a}  &         8873 \\ 
 \tableline
 Total number of sources     &       170299    \\ 
 \enddata
 \tablenotetext{a}{Including 77 recovered sources with uncertain photometry (see \S~\ref{additional_sources}). }
 \end{deluxetable}
 
 \begin{deluxetable}{lrrrr}
 \tablecaption{Statistics of sources toward VMR-D cloud (per band) \label{tab_stat_perband}}
 \tablewidth{0pt}
 \tablehead{ \colhead{Detections\tablenotemark{a} } & \colhead{3.6 $\mu$m} & \colhead{4.5 $\mu$m} & \colhead{5.8 $\mu$m} & \colhead{8.0 $\mu$m} }
 \startdata 
 S/N $\ge$10                                                 & 100964  & 81945  & 11651 & 6123   \\
 ~~~~~~ \& 24 $\mu$m                        &    764  &   801  &   634 &  603   \\
 ~~~~~~ \& 24 $\mu$m \& $^{12}$CO           &    489  &   504  &   387 &  356   \\
 ~~~~~~ \& 24 $\mu$m \& $^{12}$CO \& Dust   &    48   &    50  &    39 &   31   \\
 \tableline
 5$\le$ S/N $<$10                                            &  50291  & 48732  & 10442 & 5586   \\ 
 ~~~~~~ \& 24 $\mu$m                        &     96  &    90  &    68 &   43   \\ 
 ~~~~~~ \& 24 $\mu$m \& $^{12}$CO           &     59  &    63  &    39 &   22   \\ 
 ~~~~~~ \& 24 $\mu$m \& $^{12}$CO \& Dust   &      3  &     2  &     7 &    3   \\ 
\tableline
 Bad photometry                             &   19044 &  39622 & 148206 & 158590   \\ 
\enddata
\tablenotetext{a}{Association with a 24 $\mu$m source is within 6\arcsec, with 
 $^{12}$CO is inside the 5 K~km~s$^{-1}$ contour level, and with a dust core is 
 considered within the radii quoted in \citet{Mas07}.}
\end{deluxetable}

\begin{deluxetable}{lcccccccccc}
\tabletypesize{\scriptsize} 
\tablewidth{0pt}
\rotate
\tablecaption{YSO classification of the IRAC VMR-D sources and comparison with 
other star forming regions.\label{tab_stat_YSO}}
\tablehead{Region & Area   & Distance & YSO       & YSO          &     YSO     & Class I & flat & Class II & Class III  & SFR/Area \\ 
                  &(deg$^2$)&  (pc)   & candidate & (deg$^{-2}$) & (pc$^{-2}$) &         &      &          &         & (M$_\sun$Myr$^{-1}$
pc$^{-2}$)  \\
} 
\startdata
ON-cloud\tablenotemark{a} &  0.57 &  700  &  620 & 1088  & 7.3  & 60  (9.7\%) & 70 (11.3\%) & 247 (39.8\%) & 243 (39.2\%)  & 5.5   \\
OFF-cloud\tablenotemark{a}&  0.60 &  700  &  261 &  435  & 2.9  & 11  (4.2\%) & 18 (6.9\%)  &  99 (37.9\%) & 133 (51.0\%)  & 2.2   \\
ON-OFF\tablenotemark{b}   &  0.57 &  700  &  372 &  653  & 4.4  & 50 (13.3\%) & 53 (14.2\%) & 153 (41.1\%) & 117 (31.4\%)  & 3.3 \\
\cline{1-11}
ON-cloud\tablenotemark{c} &  0.57 &  700  &  463 &  812  & 5.4  & 44  (9.5\%) & 61 (13.2\%) & 230 (49.7\%) & 128 (27.6\%)  & 4.1   \\
OFF-cloud\tablenotemark{c}&  0.60 &  700  &  174 &  290  & 1.9  &  5  (2.9\%) & 10 (5.7\%)  &  93 (53.4\%) &  66 (37.9\%)  & 1.5   \\
ON-OFF\tablenotemark{b}   &  0.57 &  700  &  278 &  522  & 3.5  & 39 (13.2\%) & 51 (17.3\%) & 142 (47.6\%) &  65 (21.9\%)  & 2.6 \\
\cline{1-11}
ON \tablenotemark{d}      &  0.57 &  700  &  465 &  816  & 5.5  &   \nodata   &   \nodata   &   \nodata    &   \nodata     & 4.1  \\
OFF \tablenotemark{d}     &  0.60 &  700  &  196 &  326  & 2.2  &   \nodata   &    \nodata  &    \nodata   &   \nodata     & 1.6 \\
ON-OFF \tablenotemark{b}  &  0.57 &  700  &  279 &  490  & 3.3  &   \nodata   &     \nodata &    \nodata   &    \nodata    & 2.5  \\
\cline{1-11}
Cha II          & 1.04 &  178  &   26 &   25  & 2.6  &  2  (8\%) &  1 ( 4\%) &  19 (73\%) &   4 (15\%)  & 0.65\\
Lupus           & 3.10 &  150  &    9 &   30  & 3.3  &  5  (5\%) & 10 (11\%) &  52 (55\%) &  27 (29\%)  & 0.83\\
Perseus         & 3.86 &  250  &  385 &  100  & 5.2  & 87 (23\%) & 42 (11\%) & 225 (58\%) &  31  (8\%)  & 1.3 \\
Serpens         & 0.85 &  260  &  227 &  267  &  13  & 36 (16\%) & 23 (10\%) & 140 (62\%) &  28 (12\%)  & 3.2 \\
Ophiucus        & 6.60 &  125  &  292 &   44  & 9.3  & 35 (12\%) & 47 (16\%) & 176 (60\%) &  34 (12\%)  & 2.3 \\
\enddata
\tablenotetext{a}{Sources selected with the \citet{Har06,Har07} criteria.}
\tablenotetext{b}{Difference counts are weighted by the corresponding solid angles.}
\tablenotetext{c}{Contamination subtracted following \citet{Gut09,Mar08}.}
\tablenotetext{d}{Contamination subtracted according to \citet{Oli09}.}
\medskip
\end{deluxetable}

 \begin{deluxetable}{lccccccccccccccccccccccccccc}
 \tabletypesize{\footnotesize}
 \tablecaption{VMR-D IRAC candidate YSO list: typical entries\tablenotemark{a}. \label{tab_IRAC_YSO}}
 \tablewidth{0pt}
 \tablehead{ 
     \colhead{IRAC-ID} & \colhead{RA(2000)} & \colhead{Dec(2000)} &  \colhead{Glon}    & \colhead{Glat}           & \colhead{F3.6}           & \colhead{E3.6}           & \colhead{Q3.6}  \\  
                       & \colhead{F4.5}     & \colhead{E4.5}      &  \colhead{Q4.5}    &  \colhead{F5.8}          & \colhead{E5.8}           & \colhead{Q5.8}           & \colhead{F8.0}  \\  
                       & \colhead{E8.0}     & \colhead{Q8.0}      &  \colhead{MIPS ID} & \colhead{$\Delta\theta$} & \colhead{F24}            & \colhead{E24}            & \colhead{Q24} \\
		       & \colhead{F70}      & \colhead{E70}       & \colhead{Q70}      & \colhead{$\alpha$\tablenotemark{b}}  & \colhead{$^{12}$CO(1-0)} & \colhead{$^{13}$CO(2-1)} & \colhead{Dust} \\
           }
 \startdata  
   4378  & 131.224365  & -43.926624  & 263.602173 & -0.676375   & 1.02738E-03 & 2.88283E-05 &  H1 \\ 
         & 9.57767E-04 & 3.59507E-05 & H1         & 1.23024E-03 & 3.53127E-05 & H1          & 1.36355E-03 \\ 
         & 4.95515E-05 & H1          &  -         & 0.0         & 2.721E-04   & 0.000E+00   & u  \\
         & 1.199E-03   & 0.000E+00   & u          & -0.60       & N           & N           &       -    \\ 
\nodata  &  \nodata    &  \nodata    & \nodata    &  \nodata    &  \nodata    & \nodata     &  \nodata   \\ 
   38674 & 131.603989  & -43.704887  & 263.599640 & -0.323964   & 2.85320E-03 & 7.00859E-05 & H2   \\ 
         & 3.52582E-03 & 6.94163E-05 & H2         & 4.25391E-03 & 8.60992E-05 & H2          & 5.84348E-03 \\ 
         & 9.94560E-05 & H2          & 261        & 1.1         & 4.539E-02   & 8.235E-04   & G  \\
         & 9.515E-04   & 0.000E+00   & u          & 0.53        & C           & N           & umms1 \\ 
\nodata  &  \nodata    &  \nodata    & \nodata    &  \nodata    &  \nodata    & \nodata     &  \nodata   \\ 
 \enddata
\tablenotetext{a}{Symbols and units as in Table~\ref{tab_IRAC_PSC}, unless otherwise specified. Sources selected out of the working sample. }
\tablenotetext{b}{Spectral index of the source.}
 \end{deluxetable}

\begin{deluxetable}{lcccccccc} 
\tabletypesize{\footnotesize}   
\tablecaption{VMR-D MIPS candidate YSO list: typical entries.\tablenotemark{a} \label{tab_MIPS_YSO}}
\tablewidth{0pt}
\tablehead{\colhead{MIPS-ID} & \colhead{RA(2000)}        & \colhead{Dec(2000)}                        & \colhead{Glon}     & \colhead{Glat}    & \colhead{F24}              & \colhead{E24}            & \colhead{F70} \\
                             & \colhead{E70} & \colhead{$\Delta\theta$}  & \colhead{K$_s$-24\tablenotemark{b}} & \colhead{IRAC-ID} & \colhead{$^{12}$CO(1-0)} & \colhead{$^{13}$CO(2-1)} & \colhead{Dust}   \\
          }
\startdata      
  37    & 131.09596 & -43.77494  & 263.42578 & -0.65472 &  2.543E-02 & 3.492E-03  &  0.000E+00  \\ 
        & 0.000E+00 &  0.0      &  4.70      & 393       &  N	& N          &	 -   \\
\nodata & \nodata   & \nodata    & \nodata   & \nodata  &  \nodata   & \nodata    & \nodata     \\ 
 546    & 131.92834 & -43.73021  & 263.76633 & -0.15719 &  1.163E+00 & 8.047E-02  &  3.519E+00  \\ 
        & 3.890E-02 &  1.2      &  10.30     &   74548   &   C	& O          & umms11  \\
\nodata & \nodata   & \nodata    & \nodata   & \nodata  &  \nodata   & \nodata    & \nodata     \\ 
\enddata
\tablenotetext{a}{Symbols and units as in Table~\ref{tab_MIPS_cat}, unless otherwise specified.}
\tablenotetext{b}{K$_s$-24 color index in magnitudes. } 
\end{deluxetable}

 \begin{deluxetable}{cccccccccccc}
 \tabletypesize{\scriptsize}
 \tablecaption{Class~I YSOs selected with 2MASS/IRAC/MIPS data. \label{tab_IRAC_YSO_CI} }
 \tablewidth{0pt}
 \tablehead{ 
 \colhead{IRAC}    & \colhead{$\alpha$(2000)}  & \colhead{$\delta$(2000)}  & \colhead{$\alpha$} & \colhead{K$_s$} & \colhead{MIPS} & \colhead{Q24}  & \colhead{Q70}  & \colhead{MIPS}       & \colhead{$^{12}$CO} & \colhead{dust core\tablenotemark{c}} & \colhead{IRAC} \\
  \colhead{ID\tablenotemark{a}}     &  \colhead{(deg)}          & \colhead{(deg)}           &                    & \colhead{2MASS} &  \colhead{ID}  &     &     &  \colhead{class\tablenotemark{b}}          &  \colhead{(1-0)}    &                     &\colhead{variable\tablenotemark{d}} 
  }  
 \startdata 
  17793 & 131.390823  & -43.872124 & 0.61 & Y &    - &   u  &  u  &  -  &  C  &    MMS2  &  N \\
  18177 & 131.395065  & -43.882664 & 0.37 & Y &    - &   u  &  u  &  -  &  C  &       -  &  Y \\
  18652 & 131.400284  & -43.891411 & 0.82 & N &    - &   u  &  u  &  -  &  C  &       -  &  N \\
  19225 & 131.406204  & -43.917553 & 1.16 & N &    - &   u  &  u  &  -  &  C  &       -  &  N \\
  19354 & 131.407394  & -43.918594 & 0.66 & Y &    - &   b  &  u  &  -  &  C  &       -  &  N \\
  21173 & 131.426315  & -43.456017 & 0.62 & N &    - &   u  &  u  &  -  &  N  &       -  &  N \\
  21825 & 131.432907  & -43.451939 & 0.59 & N &  166 &   G  &  G  &  -  &  N  &       -  &  N \\
  22179 & 131.436584  & -43.448711 & 0.61 & Y &    - &   u  &  u  &  -  &  N  &       -  &  N \\
  22262 & 131.437500  & -43.447865 & 0.31 & Y &    - &   u  &  u  &  -  &  N  &       -  &  N \\
  23011 & 131.445084  & -43.389664 & 1.29 & N &  172 &   G  &  G  &  -  &  N  &       -  &  N \\
  38674 & 131.603989  & -43.704887 & 0.53 & Y &  261 &   G  &  u  &  I  &  C  &   umms1  &  N \\
  39589 & 131.613586  & -43.711151 & 1.25 & N &  270 &   G  &  G  &  -  &  C  &   umms1  &  N \\
  41404 & 131.632233  & -43.910416 & 1.39 & N &    - &   u  &  u  &  -  &  C  &    MMS4  &  N \\
  41429 & 131.632477  & -43.549744 & 1.04 & N &  290 &   G  &  u  &  -  &  C  &       -  &  N \\
  41786 & 131.636246  & -43.911308 & 0.89 & Y &    - &   u  &  u  &  -  &  C  &    MMS4  &  N \\
  44510 & 131.664520  & -43.380684 & 0.55 & N &  316 &   G  &  u  &  -  &  C  &       -  &  Y \\
  47368 & 131.694290  & -43.336273 & 0.54 & Y &    - &   u  &  u  &  -  &  C  &       -  &  N \\
  47436 & 131.694946  & -43.887310 & 1.37 & N &  347 &   G  &  u  &  -  &  C  &       -  &  N \\
  48046 & 131.701172  & -43.886124 & 1.08 & N &  352 &   G  &  u  &  -  &  C  &    MMS5  &  N \\
  50748 & 131.727905  & -43.876759 & 0.67 & Y &  386 &   G  &  u  &  I  &  C  &       -  &  Y \\
  55956 & 131.776840  & -43.322323 & 0.50 & N &  426 &   G  &  u  &  -  &  C  &       -  &  N \\
  65153 & 131.855423  & -43.815571 & 1.03 & N &  473 &   G  &  G  &  -  &  C  &       -  &  N \\
  65198 & 131.855774  & -43.814709 & 0.73 & Y &  473 &   G  &  G  &  -  &  C  &       -  &  N \\
  67878 & 131.877563  & -43.472893 & 0.30 & N &  497 &   G  &  u  &  -  &  C  &       -  &  Y \\
  67896 & 131.877747  & -43.467525 & 0.33 & N &  498 &   G  &  u  &  -  &  C  &       -  &  N \\
  68154 & 131.880020  & -43.898384 & 1.10 & N &  503 &   G  &  G  &  -  &  C  &       -  &  N \\
  68231 & 131.880646  & -43.896694 & 0.43 & N &  503 &   G  &  G  &  -  &  C  &       -  &  N \\
  70412 & 131.897675  & -43.356682 & 0.45 & N &  521 &   G  &  u  &  -  &  C  &       -  &  N \\
  72549 & 131.913589  & -43.438141 & 0.54 & N &    - &   u  &  u  &  -  &  C  &       -  &  N \\
  84160 & 131.999054  & -43.653702 & 0.43 & N &  601 &   G  &  u  &  -  &  C  &       -  &  N \\
 101621 & 132.129349  & -43.517223 & 0.85 & N &  717 &   G  &  u  &  -  &  C  &       -  &  N \\
 102538 & 132.136902  & -43.515751 & 0.79 & N &    - &   u  &  u  &  -  &  C  &       -  &  N \\
 104823 & 132.155701  & -43.281319 & 0.51 & N &  746 &   G  &  u  &  -  &  C  &  umms22  &  N \\
 104849 & 132.155991  & -43.280067 & 0.36 & N &  746 &   G  &  u  &  -  &  C  &  umms22  &  N \\
 106972 & 132.173294  & -43.529209 & 1.78 & N &  775 &   G  &  G  &  -  &  C  &    MMS9  &  N \\
 108306 & 132.184921  & -43.268906 & 1.44 & N &  787 &   G  &  G  &  -  &  C  &       -  &  N \\
 111144 & 132.209610  & -43.519905 & 1.83 & N &  825 &   G  &  u  &  -  &  C  &   MMS14  &  N \\
 112469 & 132.221695  & -43.515335 & 2.11 & N &  838 &   G  &  G  &  -  &  C  &   MMS16  &  N \\
 112472 & 132.221710  & -43.636738 & 0.94 & Y &  840 &   G  &  u  &  I  &  C  &       -  &  N \\
 113053 & 132.227066  & -43.553127 & 0.59 & Y &    - &   u  &  u  &  -  &  C  &       -  &  N \\
 114979 & 132.244598  & -43.639202 & 2.02 & N &  858 &   G  &  G  &  -  &  C  &   MMS17  &  N \\
 120812 & 132.302841  & -43.633789 & 0.54 & Y &    - &   b  &  u  &  -  &  C  &       -  &  N \\
 120962 & 132.304306  & -43.596691 & 0.73 & Y &    - &   u  &  u  &  -  &  C  &   MMS20  &  Y \\
 122296 & 132.318268  & -43.612213 & 0.35 & N &  947 &   G  &  u  &  -  &  C  &   MMS21  &  N \\
 124358 & 132.339371  & -44.030910 & 0.73 & N &    - &   u  &  u  &  -  &  C  &       -  &  N \\
 124521 & 132.340988  & -44.033684 & 2.07 & N &    - &   u  &  u  &  -  &  C  &       -  &  Y \\
 137019 & 132.466995  & -43.378731 & 0.46 & Y & 1108 &   G  &  u  &  I  &  C  &       -  &  N \\
 138002 & 132.477051  & -43.388687 & 0.87 & Y & 1120 &   G  &  u  &  I  &  C  &       -  &  N \\
 139510 & 132.491974  & -43.381367 & 0.88 & N & 1130 &   G  &  u  &  -  &  C  &  umms26  &  N \\
\enddata
\tablenotetext{a}{Sources selected out of the working sample.}
\tablenotetext{b}{Classification based on the diagram K$_s$ vs K$_s$-[24] (see Fig.\ref{K_24+24_70}).}
\tablenotetext{c}{Dust core names are from \citet{Mas07}.}
\tablenotetext{d}{IRAC variability on a timescale of six months \citep{Gia09}.}
\end{deluxetable}

 \begin{deluxetable}{cccccccccc}
 \tabletypesize{\scriptsize}
 \tablecaption{Class~I YSOs selected with 2MASS/MIPS data. \label{tab_MIPS_YSO_CI}}
 \tablewidth{0pt}
 
 \tablehead{ 
  \colhead{MIPS}  & \colhead{$\alpha$(2000)} & \colhead{$\delta$(2000)}  & \colhead{K$_s$-24} & \colhead{F70}  & \colhead{IRAC}        & \colhead{IRAC}  & \colhead{IRAC }                  & \colhead{$^{12}$CO} & \colhead{dust core} \\
  \colhead{ID}    & \colhead{(deg) }         & \colhead{(deg)}           &  \colhead{(mag)}   &                & \colhead{counterpart} & \colhead{flags} & \colhead{class\tablenotemark{a}} & \colhead{(1-0)}     &      \\
 }  
 \startdata 
  93 &  131.29494  & -43.75684 &  8.50 & N &  8642  & H1 H2 H1 H2 & f &  N &	-  \\
 130 &  131.38113  & -43.89581 &  8.90 & N &   16862  &  b H2  b  u & - &  N &		-  \\
 138 &  131.38982  & -43.87410 &  9.90 & N &   17713  &  b H2 H2 H2 & - &  C &  MMS2  \\
 261 &  131.60376  & -43.70510 &  9.30 & N &   38674  & H2 H2 H2 H2 & I &  C &  umms1 \\ 
 308 &  131.65639  & -43.87086 & 10.80 & Y &   43665  & H2  b  b  b & - &  C &		-  \\
 315\tablenotemark{b} &  131.66409  & -43.88627 & 8.70 & Y &	-     &             & - &  C &	   -  \\
 386 &  131.72771  & -43.87690 & 10.80  & N &	50748  & H2 H2 H2 H2 & I &  C & 	  -  \\
 541 &  131.91966  & -43.43661 &  9.20  & N &	73282  &  b H2  b H2 & - &  C & umms10 \\ 
 546 &  131.92834  & -43.73021 & 10.30  & Y &	74548  & H2 H2 H2  b & - &  C & umms11 \\ 
 576 &  131.96498  & -43.42231 &  9.50  & Y &	79572  & H2 H2 H2  b & - &  C & 	  -  \\
 729 &  132.14119  & -43.51330 &  9.50  & Y &  103109  & H2 H2  b H2 & - &  C & 	  -  \\
 814 &  132.20293  & -43.54186 &  8.40  & Y &  110380  &  b  b  b  b & - &  C &  MMS12 \\
 840 &  132.22160  & -43.63700 &  9.30  & N &  112472  & H2 H2 H2 H2 & I &  C & 	  -  \\
 884\tablenotemark{c} &  132.26060  & -44.40168 & 8.80 & N &	-     &             & - &  N &	   -  \\
 921 &  132.29868  & -43.59826 &  9.40  & N &  120425  & H2 H2 H2 H2 & f &  C &  MMS20 \\ 
 928 &  132.30367  & -43.60686 & 10.80  & Y &  120963  &  b  b H2 H2 & - &  C &  MMS21 \\ 
1108 &  132.46693  & -43.37880 &  9.30  & N &  137019  & H2 H2 H2 H2 & I &  C & 	  -  \\
1114 &  132.47073  & -43.68967 &  8.90  & N &  137393  &  b H2 H2 H2 & - &  O & 	  -  \\
1120 &  132.47690  & -43.38887 &  9.30  & N &  138002  & H2 H2 H2 H2 & I &  C & 	  -  \\
1170 &  132.54562  & -43.28630 & 10.00  & N &  144801  & H1 H2 H1  b & - &  C &  MMS29 \\ 
\enddata
\tablenotetext{a}{Classification based on the spectral index.}
\tablenotetext{b}{Embedded in very bright nebulosity near IRS17.}
\tablenotetext{c}{Observed at 24~$\mu$m only.}
\end{deluxetable}

 \begin{deluxetable}{cclccccccc}
 \tabletypesize{\scriptsize}
 \tablecaption{Other remarkable objects selected with MIPS. \label{tab_MIPS_remark}}
 \tablewidth{0pt}
 \tablehead{ 
\colhead{MIPS}                 & \colhead{$\alpha$(2000)} & \colhead{$\delta$(2000)} & \colhead{F24} &  \colhead{F70} & \colhead{IRAC}        & \colhead{IRAC}  & \colhead{Distance} & \colhead{$^{12}$CO}  & \colhead{dust core} \\
\colhead{ID\tablenotemark{a}}  &  \colhead{(deg)}         &  \colhead{(deg)}         &               &                & \colhead{counterpart} & \colhead{flags} & \colhead{(arcsec)} & \colhead{(1-0)}      &                     \\
 }  
 \startdata 
  31                   &  131.03918 &  -44.00486 &  N      &  Y &    -    &	        &     & N &      - \\
  48                   &  131.13193 &  -43.83388 &  Y      &  Y &    -    &	        & 3.4 & N &      - \\
 146\tablenotemark{b}  &  131.39792 &  -43.85200 &  S      &  Y &  18476  &  b  b  b  b & 2.8 & C &      - \\
 288                   &  131.63107 &  -43.92499 &  Y      &  Y &  41283  &  b  b  b  b & 0.5 & C &   MMS4 \\
 294                   &  131.63802 &  -43.92678 &  Y      &  Y &  41975  &  b  b  b  b & 0.9 & C &   MMS4 \\
 300\tablenotemark{b}  &  131.64540 &  -43.91031 &  S      &  Y &    -    &             & 6.5 & C &   MMS4 \\ 
 315                   &  131.66409 &  -43.88627 &  Y      &  Y &    -    &	        &16.0 & C &	 - \\
 642                   &  132.04202 &  -43.35051 &  Y      &  Y &  90122  & H2 H2 L2 L2 & 0.8 & C &      - \\
 657                   &  132.06612 &  -43.78766 &  Y      &  Y &  93485  &  b H2  b  u & 2.2 & C & umms16 \\
 661\tablenotemark{c}  &  132.06978 &  -43.78875 &  Y      &  Y &  93906  & H2 H2 H2 H2 & 0.9 & C & umms16 \\
 813\tablenotemark{b}  &  132.20084 &  -42.90561 &  S      &  Y &    -    &             & 4.9 & N &      - \\
 814\tablenotemark{b}  &  132.20293 &  -43.54186 &  S      &  Y & 110380  &  b  b  b  b & 2.1 & C &  MMS12 \\   
 917	               &  132.29530 &  -43.49447 &  Y      &  Y & 120052  & H2 H2  u  u & 3.2 & O &      - \\
 923	               &  132.30054 &  -44.27600 &  Y      &  Y &    -    &	        &11.1 & O &      - \\
 925\tablenotemark{d}  &  132.30151 &  -43.27896 &  N      &  Y & 120280  & H2 H2  u  u &14.6 & C &      - \\
 951                   &  132.32057 &  -43.93255 &  Y      &  Y & 122524  &  u  b  u L2 & 0.1 & C &      - \\
 991\tablenotemark{b}  &  132.35916 &  -43.28617 &  S      &  Y & 126339  &  b  b  b  b & 1.5 & C &  MMS22 \\ 
 998                   &  132.36937 &  -44.07478 &  Y      &  Y & 127325  &  b  b  b  b & 1.5 & C &  MMS23 \\
1018\tablenotemark{b}  &  132.38667 &  -44.18056 &  S      &  Y &    -    &             & 3.6 & C &  MMS26 \\ 
1247\tablenotemark{e}  &  132.65999 &  -43.82816 &  N      &  Y & 156273  &  u  b H2 H2 &12.0 & N &      - \\
1306	               &  132.87169 &  -43.52266 &  Y      &  Y &    -    &	        & 9.3 & N &      - \\
\enddata
\tablenotetext{a}{All the sources fall in the area mapped with IRAC and MIPS (24$\mu$m and 70$\mu$m).}
\tablenotetext{b}{Sources saturated at 24~$\mu$m: \# 146, 300, 813, 814, 991, 1018 corresponding to 
                  IRS 16, 17, 18, 19, 20, 21 of \citet{Lis92}, respectively.}
\tablenotetext{c}{Classified as Class II with IRAC.}
\tablenotetext{d}{Barely visible at 70~$\mu$m, in the other bands this position  
                  corresponds to the boundary of a spherical blob (~2\arcmin~in diameter); 
                  the IRAC source could not be a genuine association due to the large 
                  distance (column~8).} 
\tablenotetext{e}{Barely visible at 70~$\mu$m; at 24~$\mu$m there is a very strong diffuse 
                  emission and a faint object is visible at a distance of $\sim$~12\arcsec~ 
                  NW. This faint source is excluded from the 24~$\mu$m list because of the 
                  constraints imposed to the PSF fitting parameters (see \S~\ref{MIPS_phot}).}  
\end{deluxetable}

\end{document}